\documentclass[a4paper,11pt,aps,preprintnumbers,amsmath,amssymb,superscriptaddress,nofootinbib]{article}
\pdfoutput=1
\usepackage{jcappub}

\usepackage{multirow}
\usepackage{amssymb}
\usepackage{enumerate}
\usepackage{amssymb}
\usepackage{amsmath}
\usepackage{tikz}
\usepackage{feynmf}
\usepackage{makecell}
\usepackage{slashed}
\usepackage{natbib}
\usepackage{graphicx}
\hypersetup{colorlinks=true}

\usepackage{mathrsfs,amssymb}  
\usepackage{cancel}
\usepackage[normalem]{ulem}
\usepackage{slashbox}

\definecolor{darkgreen}{rgb}{0.01, 0.75, 0.24}
\definecolor{darkorange}{rgb}{1.0, 0.55, 0.0}
\usepackage{pifont}
\newcommand{\comment}[1]{}
\newcommand{\beq}{\begin{equation}}
\newcommand{\eeq}{\end{equation}}
\newcommand\bea{\begin{eqnarray}}
\newcommand\eea{\end{eqnarray}}

\newcommand{\eq}[1]{Eq.~(\ref{#1})}
\newcommand{\eqs}[2]{Eqs.~(\ref{#1}) and (\ref{#2})}

\newcommand{\TeV}{\mathinner{\mathrm{TeV}}}

\renewcommand{\l}{\left}
\renewcommand{\r}{\right}

\title{Cosmology with a supersymmetric local $B-L$ model}

\author{Kwang Sik Jeong$^1$}
\author{and Wan-il Park$^{2,3,4}$}
\affiliation{
$^1$Department of Physics, Pusan National University, Busan 46241, Korea\\
$^2$Division of Science Education and Institute of Fusion Science, Jeonbuk National University, Jeonju 54896, Republic of Korea\\
$^3$Instituto de Física Corpuscular, CSIC-Universitat de València, Paterna 46980, Spain\\
$^4$Departament de Física Teòrica, Universitat de València, Burjassot 46100, Spain	
}

\abstract{
We propose a minimal gauged $U(1)_{B-L}$ extension of the minimal supersymmetric Standard Model (MSSM) which resolves the cosmological moduli problem via thermal inflation, and realizes late-time Affleck-Dine leptogensis so as to generate the right amount of baryon asymmetry at the end of thermal inflation.
The present relic density of dark matter can be explained by sneutrinos, MSSM neutralinos,  axinos, or axions.
Cosmic strings from $U(1)_{B-L}$ breaking are very thick, and so the expected stochastic gravitational wave background from cosmic string loops has a spectrum different from the one in the conventional Abelian-Higgs model, as would be distinguishable at least at LISA and DECIGO.
The characteristic spectrum is due to a flat potential, and may be regarded as a hint of supersymmetry.
Combined with the resolution of moduli problem, the expected signal of gravitational waves constrains the $U(1)_{B-L}$ breaking scale to be $\mathcal{O}(10^{12-13})\,{\rm GeV}$.
Interestingly,   our model provides a natural possibility for explaining the observed ultra-high-energy cosmic rays thanks to the fact that the core width of strings in our scenario is very large, allowing a large enhancement of particle emissions from the cusps of string loops.
Condensation of $LH_u$ flat-direction inside of string cores arises inevitably and can also be the main source of the ultra-high-energy cosmic rays 
accompanied by ultra-high-energy lightest supersymmetric particles. 
}

\emailAdd{ksjeong@pusan.ac.kr}
\emailAdd{wipark@jbnu.ac.kr}

\begin{document}

\maketitle

\section{Introduction}
\label{sec:intro}

The standard model (SM) of particle physics has been really successful in describing a sub-atomic world.
However, it suffers from the so-called hierarchy problem between the electroweak and Planck scales. 
Also, there are many experimental and observational evidences,  such as neutrino flavor oscillations, dark matter, 
baryon asymmetry in the universe,   hinting for physics  beyond the SM (BSM).

As the largest possible spacetime symmetry
and  a very well-motivated framework for BSM theories, 
supersymmetry (SUSY)
provides a natural solution to the hierarchy problem if the mass scale of superpartners of the SM particles is not very higher than  
the electroweak scale.
Also, in the minimal supersymmetric standard model (MSSM) with such a SUSY  scale, 
the SM gauge couplings unify precisely without heavy threshold corrections at a scale around the grand unification scale, 
$M_{\rm  GUT} \sim 10^{16} \,{\rm GeV}$.
Moreover, if $R$-parity is conserved, the MSSM possesses a plausible dark matter candidate,
the lightest supersymmetric particle (LSP),  which can naturally 
have the right amount of relic density in the present universe.
However,  including the discovery of the SM Higgs boson at $125\, {\rm GeV}$,  
the absence of SUSY signals in various searches for the lightest neutralino dark matter has been pushing up 
the SUSY  scale at least to $\mathcal{O}(1-10) {\rm TeV}$ scale~\cite{Canepa:2019hph,ATLAS:2023summary,CMS:2023qhl}.
This implies that, even though SUSY removes the big hierarchy between the SUSY and Planck scales,
we face to a little hierarchy problem between the electroweak and SUSY scales.
Also,  in high scale SUSY, the so-called WIMP miracle does not work any more,  
and LSP neutralinos are very likely to be over-produced except in some special region of parameter space. 

Meanwhile, in the framework of supergravity,  which is believed to be the low energy effective theory of string theory,
moduli (Planckian flat directions) are ubiquitous.
If their masses are of $\mathcal{O}(1-10)\,{\rm TeV}$ scale, 
they decay during the epoch of big bang nucleosynthesis (BBN), invalidating the theory of BBN that is in a  good agreement 
with observations.
Pushing up the SUSY scale  is good for remedying this cosmological moduli problem, but the moduli-induced gravitino problem still 
remains even in models with a light sparticle, 
such as the axino,  into which the lightest neutralino decays~\cite{Endo:2006zj}.

Thermal inflation provides the most compelling solution to the cosmological moduli problem~\cite{Lyth:1995hj,Lyth:1995ka}, 
owing to the late-time production of a huge amount of entropy after its end.
In order to recover the standard thermal bath before BBN, the flaton responsible for the end of thermal inflation should decay 
dominantly into SM particles.
Although there might be possibilities of cascade decays via some unknown sectors, the conventional way of recovering SM thermal bath is to couple the flaton to the Higgs bilinear operator.
Such a coupling can generate the MSSM Higgs bilinear $\mu$-term once the flaton develops an intermediate vacuum expectation value.
It is important to note that 
the dynamical generation of $\mu$-term can trigger a late-time Affleck-Dine leptogenesis~\cite{Stewart:1996ai,Jeong:2004hy,Kim:2008yu,Park:2010qd}.
It is otherwise non-trivial to generate the right amount of baryon asymmetry in the presence of thermal inflation.

The flaton associated with thermal inflation might be identified as the Peccei-Quinn (PQ) field which implements the axion solution to the strong CP problem~\cite{Kim:1986ax,Peccei:2006as}.
However, in such a case, the PQ breaking scale  is limited to be less than $\mathcal{O}(10^{10})\, {\rm GeV}$ because  PQ   breaking occurs after primordial inflation and produces axions dominantly via the decay of unstable string-wall networks~\cite{Kawasaki:2014sqa}.
The decay temperature of flaton is then likely to be too high to resolve the cosmological moduli problem, and causes too much dark matter even in the case with axino LSP~\cite{Kim:2008yu}.
 Multiple stages of thermal inflation can help resolving the moduli problem,  but still the decay temperature of the last thermal inflation should be low enough to avoid dark matter overproduction.
 
In this work, we consider a minimal local $U(1)_{B-L}$ extension of the MSSM in which the $B-L$ breaking field is identified as the flaton associated with thermal inflation, 
and is responsible for the dynamical generation of the MSSM $\mu$-term~\footnote{For earlier works in this direction, 
see Ref.~\cite{Jeannerot:1998qm} for instance.}.
In this scenario, the symmetry-breaking scale of $U(1)_{B-L}$ can be large enough such that the cosmological moduli problem is easily resolved even with the single 
stage of thermal inflation.
The right amount of baryon asymmetry of the universe is generated  by the late-time Affleck-Dine mechanism,  
and  also the observed dark matter density can be obtained from sneutrinos,  electroweak neutalinos,  or 
KSVZ-type axinos with axions.
Furthermore, 
cosmic strings formed at the $U(1)_{B-L}$-breaking at the end of thermal inflation are expected to produce a stochastic gravitational 
waves background (SGWB), which are within the reach of the near-future experiments, such as SKA~\cite{SKA}, LISA~\cite{LISA}, and DECIGO~\cite{DECIGO}.
Thanks to the flatness of the symmetry-breaking potential, the SGWB has a characteristic spectrum  distinguishable from the one expected in the conventional Abelian-Higgs model.
Interestingly,  because their core width is very large, 
cosmic strings can source the observed ultra-high-energy cosmic rays (UHECR's) over the GZK limit \cite{AlvesBatista:2019tlv}.
Condensation along the $LH_u$ flat direction arises inevitably at the inside of string cores, and can also be the origin of those UHECR's accompanied by UHE LSP's.

This paper is organized as follows.
In section 2, a local $U(1)_{B-L}$ extension of the MSSM is introduced.
In section 3, properties of thermal inflation in our scenario are discussed.
In section 4, the generation of baryon  asymmetry is briefly sketched.
In section 5,  the relic density of dark matter is estimated.
In section 6,  gravitational waves from cosmic strings loops are discussed.
In section 7,  production of UHECR's is discussed briefly with an estimation of the expected flux as an illustration.
In section 8, conclusions are drawn.

\section{The Model}
\label{sec:model}

For the gauged $U(1)_{B-L}$ scenario,   anomaly cancellation condition requires at least two
$B-L$ Higgs chiral superfields.
Here,  as a minimal setup,  we include two Higgs chiral superfields,  $\Phi_1$ and $\Phi_2$,  assuming that
they carry only 
$U(1)_{B-L}$ charges assigned as
\beq
q_{BL} (\Phi_1, \Phi_2) = (1, -1).
\eeq
Then, under the gauge group $G_{\rm SM} \otimes U(1)_{B-L}$ with $G_{\rm SM}$ being the SM gauge group, we can consider the following superpotential
\beq \label{eq:W-full}
W 
= W_{{\rm MSSM}-\mu} + \mu_H H_uH_d + \mu_\Phi \Phi_1 \Phi_2
+ \frac{1}{2} y_N \Phi_1 N^2 + y_\nu LH_u N + \Delta W_{\rm high},
\eeq 
where $W_{{\rm MSSM}-\mu}$ includes the conventional MSSM Yukawa terms,
and $\Delta W_{\rm high}$ consists of the leading gauge-invariant higher order terms 
involving  
the MSSM and $B-L$ Higgs chiral superfields only:
\beq \label{eq:W-high}
\Delta W_{\rm high} =  \frac{\lambda_H}{2M} \l( H_u H_d \r)^2 + \frac{\lambda_\mu}{M} \Phi_1 \Phi_2 H_u H_d +  \frac{\lambda_\Phi}{2M} \l( \Phi_1 \Phi_2 \r)^2,
\eeq 
where $M$ is the cutoff scale of the effective interactions,
  which may be the Planck scale.

The dimensionful parameters, $\mu_H$ and $\mu_\Phi$, of the Higgs bilinear terms, $H_u H_d$ and $\Phi_1 \Phi_2$ respectively, may not be directly from the dominant SUSY-breaking sector responsible for the soft SUSY-breaking parameters.
They might be originally absent in the renormalizable superpotential due to superconformal symmetry, but might be generated from K\"{a}hler potential, for example, via the Jiudice-Masiero mechanism \cite{Giudice:1988yz} or higher order terms.
Here we assume that the bare  values of
$\mu_H$ and $\mu_\Phi$ are much smaller than the soft SUSY-breaking scale, $m_s$.
Such smallness of the Higgs bilinear couplings may be a result of an approximate
discrete symmetry,  $Z_4$,  under which $H_uH_d\to -H_uH_d$ and $\Phi_1 \Phi_2 \to -\Phi_1\Phi_2$.
Note that $\Phi_1$ and $\Phi_2$ should develop non-zero VEVs before the electroweak symmetry breaking such that the MSSM $\mu$-term 
at low energy around or below the electroweak scale should be given by
\beq
|\mu| \equiv \l| \mu_H + \frac{\lambda_\mu v_1 v_2}{2M} \r| \lesssim m_s,
\eeq
where $v_i = \sqrt{2} \langle \Phi_i \rangle$, the VEV of the canonically normalized scalar field associated with $\Phi_i$. 
Also, we take the coupling constants in $\Delta W_{\rm high}$ as free parameters which can be much smaller than unity by several orders of magnitude.

The presence of two $U(1)$ gauge groups in general causes mixings of kinetic terms and gauge 
couplings~\cite{OLeary:2011vlq}.
We ignore those mixings for simplicity because they do not affect the main points of our argument.

\subsection{Tree-level scalar potential of 
$B-L$ Higgs fields 
at $T=0$}

The scalar potential involving the SM Higgs scalar fields,  $H_u$ and $H_d$,
and $B-L$ Higgs scalar fields,  $\Phi_1$ and $\Phi_2$,  is written as
\beq
V = V_0 + V_{\rm soft} + V_F + V_D,
\eeq
where   $V_0$ is the constant term determined by the condition of vanishing cosmological constant
in the present universe, 
and other terms are given by
\bea
V_{\rm soft} 
&=& m_{H_u}^2 |H_u|^2 + m_{H_d}^2 |H_d|^2 
- \l[ B_H \mu_H H_uH_d + \frac{A_H \lambda_H}{2M} \l( H_uH_d \r)^2 + {\rm h.c.} \r]
\nonumber \\
&+& m_1^2 |\Phi_1|^2 + m_2^2 |\Phi_2|^2
- \l[ B_\Phi \mu_\Phi \Phi_1 \Phi_2 + \frac{A_\Phi \lambda_\Phi}{2M} \l( \Phi_1 \Phi_2 \r)^2 + {\rm h.c.} \r]
\nonumber \\
&-& \l[ \frac{A_\mu \lambda_\mu}{M} H_uH_d\Phi_1\Phi_2 + {\rm h.c.} \r],
\\
V_F
&=& \l| \mu_H + \frac{\lambda_H H_uH_d}{M} + \frac{\lambda_\mu \Phi_1 \Phi_2}{M} \r|^2 \l( |H_u|^2 + |H_d|^2 \r)
\nonumber \\
&+& \l| \mu_\Phi + \frac{\lambda_\Phi \Phi_1 \Phi_2}{M} + \frac{\lambda_\mu H_uH_d}{M} \r|^2 \l( |\Phi_1|^2 + |\Phi_2|^2 \r),
\\
V_D
&=& \frac{1}{2} \l( g_Y^2 Q_Y^2 + g_2^2 Q_2^2 \r) \l( |H_u|^2 - |H_d|^2 \r)^2 
+ \frac{1}{2} g_{BL}^2 Q_{BL}^2 \l( |\Phi_1|^2 - |\Phi_2|^2 \r)^2,
\eea
with $Q_Y = Q_2 = 1/2$ and $Q_{BL}=1$. 

In the early universe with temperature well above the electroweak scale,   the electroweak symmetry can be 
recovered  to give $\langle H_u\rangle=\langle H_d \rangle = 0$.
The scalar potential is then
\bea
V
&=& V_0 + m_1^2 |\Phi_1|^2 + m_2^2 |\Phi_2|^2
- \l[ B_\Phi \mu_\Phi \Phi_1 \Phi_2 + \frac{A_\Phi \lambda_\Phi}{2M} \l( \Phi_1 \Phi_2 \r)^2 + {\rm c.c.} \r]
\nonumber \\
&+& \l| \mu_\Phi + \frac{\lambda_\Phi \Phi_1 \Phi_2}{M} \r|^2 \l( |\Phi_1|^2 + |\Phi_2|^2 \r)
+ \frac{1}{2} g_{BL}^2 Q_{BL}^2 \l( |\Phi_1|^2 - |\Phi_2|^2 \r)^2,
\eea
omitting the thermal potential terms associated with $\Phi_1$ and $\Phi_2$.
The soft scalar mass-squared ($m_{1,2}$) of the $B-L$ Higgs fields can have a negative value at low energy scales 
due to strong radiative corrections from Yukawa interactions with $y_N = \mathcal{O}(1)$ or negative mass-squared parameters at the mediation scale of supersymmetry breaking.  
We assume that this is the case and 
\beq
m_{1,2}^2 + |\mu_\Phi|^2 < 0
\eeq
from the scale of $m_s$ up to at least some intermediate scales.
In this case, $\Phi_{1,2}$ can develop non-zero VEVs at temperature at the scale of $\sqrt{| m_{1,2}^2|}$ or higher.

If stabilized in the field range much larger than $m_s$,  
$\Phi_1$ and $\Phi_2$  are located 
close to 
the $D$-flat direction,  $\Phi_1=\Phi_2 \equiv \phi/\sqrt2$,
where we have assumed for simplicity that the $B-L$ sector preserves $CP$
so that  one can take a field basis in which 
$\arg(\Phi_1)=\arg(\Phi_2)=0$ at the minimum.  
The deviation from the $D$-flat direction turns out to be (see Appendix~\ref{sec:app-dev-flat})
\beq
|\Delta \phi |\equiv \l| \Phi_2 - \Phi_1 \r| \simeq \frac{1}{2 \sqrt{2} g_{BL}^2} \frac{|\Delta m^2|}{|\phi|} ,
\eeq
where $\Delta m^2 \equiv m_2^2-m_1^2$.
Ignoring the deviation,  which is  negligibly small even relative to $m_s$,   
one finds 
\beq \label{eq:V-Dflat}
V
= V_0 + \frac{1}{2} \l( m_1^2 + m_2^2 \r) |\phi|^2
- \frac{1}{2} \l[ B_\Phi \mu_\Phi \phi^2 + \frac{A_\Phi \lambda_\Phi}{4M} \phi^4 + {\rm c.c.} \r] + \l| \mu_\Phi 
+ \frac{\lambda_\Phi \phi^2}{2 M} \r|^2 |\phi|^2.
\eeq
Under the assumption that the $B-L$ sector preserves $CP$, 
it is straightforward to see that the value of $\phi$ at the minimum,  $\phi_0 \equiv \langle \phi \rangle$,  reads
\bea \label{eq:phi0}
 \phi^2_0  
&=& \frac{A_\Phi M}{3 \lambda_\Phi} \l[ 1 - \frac{4 |\mu_\Phi|}{A_\Phi} + \sqrt{1 + \frac{12 \overline{m^2} +4 |\mu_\Phi|^2 - 8 A_\Phi \mu_\Phi + 12 B_\Phi \mu_\Phi}{A_\Phi^2}} \r]
\nonumber \\
&\approx& \frac{A_\Phi M}{3 \lambda_\Phi} \l[ 1 + \sqrt{1 + \frac{12 \overline{m^2}}{A_\Phi^2}} \r],
\eea
where $\overline{m^2} \equiv - \l(m_1^2 + m_2^2 \r)/2 > 0$,   and we have ignored the contributions involving $\mu_\Phi$ 
to estimate the approximated value.   
For $\mu_\phi \ll \sqrt{\overline{m^2}} \sim A_\Phi \sim B_\Phi$, \eq{eq:phi0} gives
\beq
V_0 \approx \overline{m^2} \phi_0^2
\eeq

\subsection{Particle spectrum and decay rates of
$B-L$ Higgs fields
}

The $B-L$ extension under consideration 
involves three new superfields,  
the two $B-L$ Higgs fields 
and the $B-L$ gauge field,  
 in addition to the MSSM and RHN-sector superfields. 
There are several light particles relevant for our discussion.
Simply speaking,   they are particles associated with the superfield of the $D$-flat direction.
The radial and phase components of the flaton field $\phi$ correspond to the physical modes under the assumption of $CP$ conservation in the $B-L$ sector.
The mass of the radial component reads
\bea
m_\phi^2 &\approx& 
\frac{1}{3} A_\Phi^2 \l( 
1 + \frac{12 \overline{m^2}}{A_\Phi^2} + \sqrt{1 + \frac{12 \overline{m^2}}{A_\Phi^2}} \r),
\eea
while that of  the phase component is given by  
\bea
m_a^2 &\approx& \frac{2}{3} A_\Phi^2 \l( 1 + \sqrt{1 + \frac{12 \overline{m^2}}{A_\Phi^2}} \r),
\eea
as follows from  \eq{eq:V-Dflat}.

If we assume for simplicity that the mixings associated with gauge-kinetic term, gauge couplings, and gauginos of $U(1)_Y$ and $U(1)_{B-L}$ are absent, then  the fermionic superpartner of the $D$-flat direction, $\psi$, is given as a linear combination of $\psi_1$ and $\psi_2$, 
\beq
\psi = \frac{1}{\sqrt{2}} \psi_1 + \frac{1}{\sqrt{2}} \psi_2 . 
\eeq
The mass of $\psi$ can be directly read out from the superpotential in \eqs{eq:W-full}{eq:W-high}, and found as
\beq
m_\psi = \mu_\Phi + \frac{3\lambda |\phi_0|^2}{2M}.
\eeq
Note that, from the tadpole condition along the $D$-flat direction, one can see that, as long as $\mu_\Phi$ is relatively small compared to $m_s$, 
the mass of $\psi$ is bounded as
\beq
m_\psi \gtrsim m_\phi.
\eeq
Hence the flaton cannot decay to its superpartner.

%

If kinematically allowed, the flaton  $\phi$ decays into SM particles and electroweak neutralinos. 
Direct decays of $\phi$ to superparticles 
are
 likely to cause over-production of LSPs unless the branching fraction to those particles is highly suppressed.
So, we assume that the mass of $\phi$ at the true vacuum is such that $\phi$ cannot decay to MSSM neutralino LSPs.

For $m_\phi$ much larger than the mass of SM Higgs, $m_h = 125 {\rm GeV}$, but smaller than the mass of the neutralino LSP, 
the decay 
of $\phi$ is dominated by the SM-Higgs channel \cite{Kim:2008yu}, 
\beq \label{eq:Gamma-phi}
\Gamma^\phi_{hh} 
= \frac{1}{4 \pi} \frac{m_\phi^3}{\phi_0^2} \l( \frac{m_A^2 - |B|^2}{m_A^2} \r)^2 \l( \frac{|\mu|}{m_\phi} \r)^4
\sim \frac{1}{4 \pi} \frac{m_\phi^3}{\phi_0^2} \l( \frac{|\mu|}{m_\phi} \r)^4,
\eeq
where  
the mass of the heavy CP-odd Higgs boson is given by 
\beq
m_A^2 \equiv m_{H_u}^2 + m_{H_d}^2 + 2 |\mu|^2,
\eeq
and $B$ denotes the soft-SUSY breaking parameter associated with the $\lambda_\mu$-term.
Electroweak symmetry breaking requires \cite{Martin:1997ns}
\beq
2 |B \mu| < m_A^2, \quad 
| B \mu |^2 > \l( |\mu|^2 - m_{H_u}^2 \r) \l( |\mu|^2 + m_{H_d}^2 \r).
\eeq
Since the minimization condition leads to 
\beq
\frac{m^2_Z}{2} 
= -|\mu|^2
+ \frac{m^2_{H_d} - m^2_{H_u} \tan^2\beta}{\tan^2\beta -1 },
\eeq
with $\tan \beta \equiv  \langle|H^0_u| \rangle /\langle |H^0_d| \rangle$,  one finds
\beq \label{eq:rel-for-mZ}
\frac{m^2_Z}{2} \simeq -|\mu|^2 -m^2_{H_u} + \frac{m^2_{H_d}}{\tan^2\beta},
\eeq
for moderate and large $\tan\beta$.
\eq{eq:rel-for-mZ} then implies
\beq
m_{H_u}^2 \approx -|\mu|^2,
\eeq
as long as $m^2_{H_d}$ is not much larger than  $m^2_{H_u}$ in size around the electroweak scale, as would be naturally the case.

\section{Thermal inflation}


Thermal inflation, which was proposed to resolve the cosmological moduli problem, is realized in our scenario as discussed already in Ref.~\cite{Jeannerot:1998qm}.
It begins when $V(\phi, T)$  at $\phi=0$ dominates the energy density of the universe while the flaton $\phi$ is fixed 
in the vicinity of the origin thanks to large thermal corrections making its effective mass squared large positive.
In the simplest case,  i.e.,  for a single stage of thermal inflation, 
the $e$-folding number of thermal inflation is given by
\beq
N_e^{\rm TI} = \ln \l( \frac{T_{\rm b}}{T_{\rm c}} \r),
\eeq
where $T_{\rm b}$ and $T_{\rm c}$ are respectively the temperature at the beginning and end of thermal inflation:
\bea
T_{\rm b} &\sim& \frac{\l( m_\phi \phi_0 \r)^{2/3}}{\l( m_\varphi M_{\rm P} \r)^{1/6}},
\\
T_{\rm c} &\sim& m_\phi / \lambda_{\rm eff},
\eea
where $m_\varphi$ is the mass of modulus,  $\varphi$, 
and $\lambda_{\rm eff}$ is the effective coupling of $\phi$ to thermal bath.
A realization of thermal inflation implies 
 $T_{\rm c} < T_{\rm b}$, which requires
\beq
\lambda_{\rm eff} \gtrsim \l( \frac{m_\varphi}{m_\phi} \r)^{1/3} \l( \frac{m_\phi}{\phi_0} \r)^{2/3} \l( \frac{M_{\rm P}}{m_\varphi} \r)^{1/6}.
\eeq
This condition is easily satisfied, and does not impose any stringent constraint 
on the model  as long as $m_\varphi \sim m_\phi$ and $\phi_0$ is of an intermediate scale.

Once thermal inflation ends, 
there can be a period of matter-domination depending on the decay rate of flaton.
The decay of flaton releases a large amount of entropy, diluting pre-existing particles.
The dilution factor is given by
\beq
\Delta_{\rm TI} \simeq \frac{V_{\rm TI}}{T_{\rm c}^3 T_{\rm d}},
\eeq
where $V_{\rm TI} \simeq V_0$ is the potential energy at $\phi=0$ that drives thermal inflation.
The decay temperature of flaton after thermal inflation can be defined such that
\beq
\rho_{\rm r}(T_{\rm d}) = \rho_\phi(T_{\rm d}).
\eeq
It gives 
\beq
H_{\rm d} = \frac{2}{5} \Gamma_\phi,
\eeq
and then one finds 
\beq \label{eq:Td}
T_{\rm d} = \l( 2 \alpha_{\rm d} \r)^{-1/4} \sqrt{\frac{2}{5} \Gamma_\phi M_{\rm P}} \sim \sqrt{\Gamma_\phi M_{\rm P}},
\eeq
where $\alpha_i \equiv \alpha(T_i) \equiv  \pi^2 g_*(T_i)/90 $,  and $\Gamma_\phi$ is the total decay rate of $\phi$.
For $m_\phi \gg m_h$,  if $\mu \sim m_s$ and $\l( m_A^2 - |B|^2 \r)/m_A^2 = \mathcal{O}(1)$, 
one finds $\Gamma_\phi \simeq \Gamma_{hh}^\phi$ \cite{Kim:2008yu}.
In terms of the decay temperature $T_{\rm d}$,  the expansion rate after the end of thermal inflation 
and before the flaton decay is written as
\beq \label{eq:H-phi-dom}
H(T) 
= \l( 1 + r \r) \sqrt{\alpha_{\rm d}/2} \l( \frac{g_*(T)}{g_*(T_{\rm d})} \r) \frac{T^4}{T_{\rm d}^2 M_{\rm P}} 
= \frac{1+r}{5} \l( \frac{g_*(T)}{g_*(T_{\rm d})} \r) \frac{T^4}{T_{\rm d}^4} \Gamma_\phi,
\eeq
with $r \equiv \rho_{\rm r}(T)/\rho_\phi(T)$.
Note that $r=1$ at $T=T_{\rm d}$.

Cosmological moduli can be produced before and after the end of thermal inflation.
The former and latter may be called as `big-bang moduli' and `thermal inflation moduli', respectively \cite{Lyth:1995ka}.
In order not to disturb the standard BBN processes, the abundance of moduli from both contributions is constrained as \cite{Kawasaki:2017bqm}
\beq \label{eq:Y-moduli-const}
Y_\varphi \lesssim \l\{
\begin{array}{ll}
10^{-17} \l( \frac{1 {\rm TeV}}{m_\varphi} \r) &: \  10^3 {\rm s} \lesssim \tau_\varphi 
\\
10^{-13} \l( \frac{10 {\rm TeV}}{m_\varphi} \r)  \sim 10^{-17} \l( \frac{1 {\rm TeV}}{m_\varphi} \r)  &: \  10 {\rm s} \lesssim \tau_\varphi \lesssim 100 {\rm s}
\\
10^{-13} \l( \frac{100 {\rm TeV}}{m_\varphi} \r) &: \  \tau_\varphi \sim 1 {\rm s}
\end{array}
\r.
\eeq
with $\tau_\varphi$ being the lifetime of the modulus $\varphi$.
The decay rate of moduli was taken to be
\beq
\Gamma_\varphi = \frac{N_{\rm ch} m_\varphi^3}{32 \pi M_{\rm P}^2},
\eeq
with $N_{\rm ch} = \mathcal{O}(10)$ being the effective number of decay channels.  
For the purpose of numerical analysis, we parametrize the above bound as
\bea
Y_\varphi^{\rm bnd} &=& 4 \times 10^{-18} \times \left\{ 
\l( \frac{1.5 \TeV}{m_\varphi} \r) \Theta(1.5 {\rm TeV} - m_\varphi) \r.
\nonumber \\
&&
\l. + \,{\rm Exp}\l[ 0.148 \times \l( \frac{m_\varphi}{1 \TeV} - 1.5 \r)^2 \r] \l( 1 - \Theta  \l( 1.5 \TeV - m_\varphi \r) \r) \Theta \l( 9.5 \TeV - m_\varphi \r) \r.
\nonumber \\
&&
\left. +\, {\rm Exp} \l[ 0.148 \times 64 \r] \l( 1 - \Theta  \l( 1.5 \TeV - m_\varphi \r) \r) \l( 1 - \Theta \l( 9.5 \TeV - m_\varphi \r) \r)
\right\},
\eea
in the range of $1 \TeV \lesssim m_\varphi \lesssim 100 \TeV$, 
where $\Theta$ is the Heaviside step function.

Even though heavy moduli can decay before the beginning of BBN, they generally decay to gravitinos with a sizable branching fraction, causing over-production of electroweak neutralinos if the mass of neutralino is larger than several hundred GeV.
This is the so-called moduli-induced gravitino problem~\cite{Endo:2006zj}.
To resolve this problem,  one may consider much lighter axino LSPs, which can  naturally arise in the case of a radiatively stabilized PQ field~\cite{Kim:2008yu}.
It requires a GeV or sub-GeV scale axino, which is however unlikely to be achieved for $m_s \gtrsim \mathcal{O}(1) {\rm TeV}$ 
\footnote{If the mass of the lightest modulus is such that its decay to gravitinos is forbidden, 
the problem associated with gravitinos can be alleviated though~\cite{Bae:2021rmg}.}. 
Hence, even in the case of heavy moduli decaying well after the freeze-out of neutralinos, thermal inflation 
would be the best solution so far to the problems associated with those moduli.

The abundance of big-bang moduli at the end of thermal inflation depends on the reheating temperature, $T_{\rm R}$, after the primordial inflation.
  For thermal inflation to occur,  one needs  
\beq
T_{\rm R} \gtrsim T_{\rm c}^2 / V_{\rm TI}^{1/4} \sim m_\phi \l( \frac{m_\phi}{\phi_0} \r)^{1/2}.
\eeq
Then,  the abundance of big-bang moduli is bounded as
\beq \label{eq:Y-varphi-BB0}
\frac{T_{\rm R}}{m_\varphi} \lesssim Y_{\varphi}^{\rm BB, 0} \lesssim \frac{\alpha_{\rm osc}^{-1/4}}{4} \l( \frac{M_{\rm P}}{m_\varphi} \r)^{1/2} \l( \frac{\varphi_0}{M_{\rm P}} \r)^2,
\eeq
where we assumed $\rho_\varphi^0 \sim m_\varphi^2 \varphi_0^2$ with $\varphi_0 \sim M_{\rm P}$ being the initial oscillation amplitude of the modulus at the time when the Hubble expansion rate becomes $H = m_\varphi$.
Here we consider arbitrarily high $T_{\rm R}$
to see if thermal inflation can completely resolve the moduli problem. 
The late time abundance of moduli before their decay is then given by
\bea \label{eq:Y-moduli-BB}
Y_\varphi^{\rm BB} 
&=& \frac{Y_\varphi^{\rm BB,0}}{\Delta_{\rm TI} }
= \frac{\alpha_{\rm osc}^{-1/4}}{4} \l( \frac{M_{\rm P}}{m_\varphi} \r)^{1/2} \l( \frac{\varphi_0}{M_{\rm P}} \r)^2 \frac{T_{\rm c}^3 T_{\rm d}}{V_{\rm TI}}
\nonumber \\
&=& \frac{\l( 2 \alpha_{\rm osc} \alpha_{\rm d} \r)^{-1/4}}{8 \sqrt{5 \pi/2} \beta^2}  \l( \frac{\varphi_0}{M_{\rm P}} \r)^2 \l( \frac{M_{\rm P}}{\sqrt{m_\varphi m_\phi}} \r) \l( \frac{\mu}{m_\phi} \r)^2 \l( \frac{m_A^2 - |B|^2}{m_A^2} \r) \l( \frac{T_{\rm c}}{\phi_0} \r)^3 ,
\eea
where we have used $V_{\rm TI} = \beta^2 m^2_\phi  \phi_0^2$ with $\beta $ being a numerical constant of order unity. 

At the end of thermal inflation, even in the case the abundance of big-bang moduli diminished, coherent oscillations of moduli are expected to reappear with an amplitude roughly given by \cite{Lyth:1995ka}
\beq
\varphi^{\rm osc} \sim \frac{V_{\rm TI}}{m_\varphi^2 M_{\rm P}^2} \varphi_0.
\eeq
The abundance of thermal inflation moduli at the end of thermal inflation is then expected to be
\beq
Y_\varphi^{\rm TI,0}  \sim \frac{\alpha_{\rm c}^{-1}}{4} \frac{\varphi_0^2 V_{\rm TI}^2}{T_{\rm c}^3 m_\varphi^3 M_{\rm P}^4},
\eeq
and the late time abundance of this type of moduli before their decay is 
\bea \label{eq:Y-moduli-TI}
Y_\varphi^{\rm TI}  &=& Y_\varphi^{\rm TI,0} /\Delta_{\rm TI} 
\sim \frac{\alpha_{\rm c}^{-1}}{4} \frac{T_{\rm d} \varphi_0^2 V_{\rm TI}}{m_\varphi^3 M_{\rm P}^4}
\\
&=& \frac{\beta^2 \alpha_{\rm c}^{-1} \l( 2 \alpha_{\rm d} \r)^{-\frac{1}{4}}}{8 \sqrt{5 \pi/2}} \l( \frac{\mu}{m_\phi} \r)^2 \l( 1-  \frac{|B|^2}{m_A^2} \r) \l( \frac{m_\phi}{m_\varphi} \r)^3 \l( \frac{m_\phi}{M_{\rm P}} \r)^{\frac{1}{2}} \l( \frac{\phi_0}{M_{\rm P}} \r) \l( \frac{\varphi_0}{M_{\rm P}} \r)^2.
\eea

\begin{figure}[h] 
\begin{center}
\includegraphics[width=0.45\textwidth]{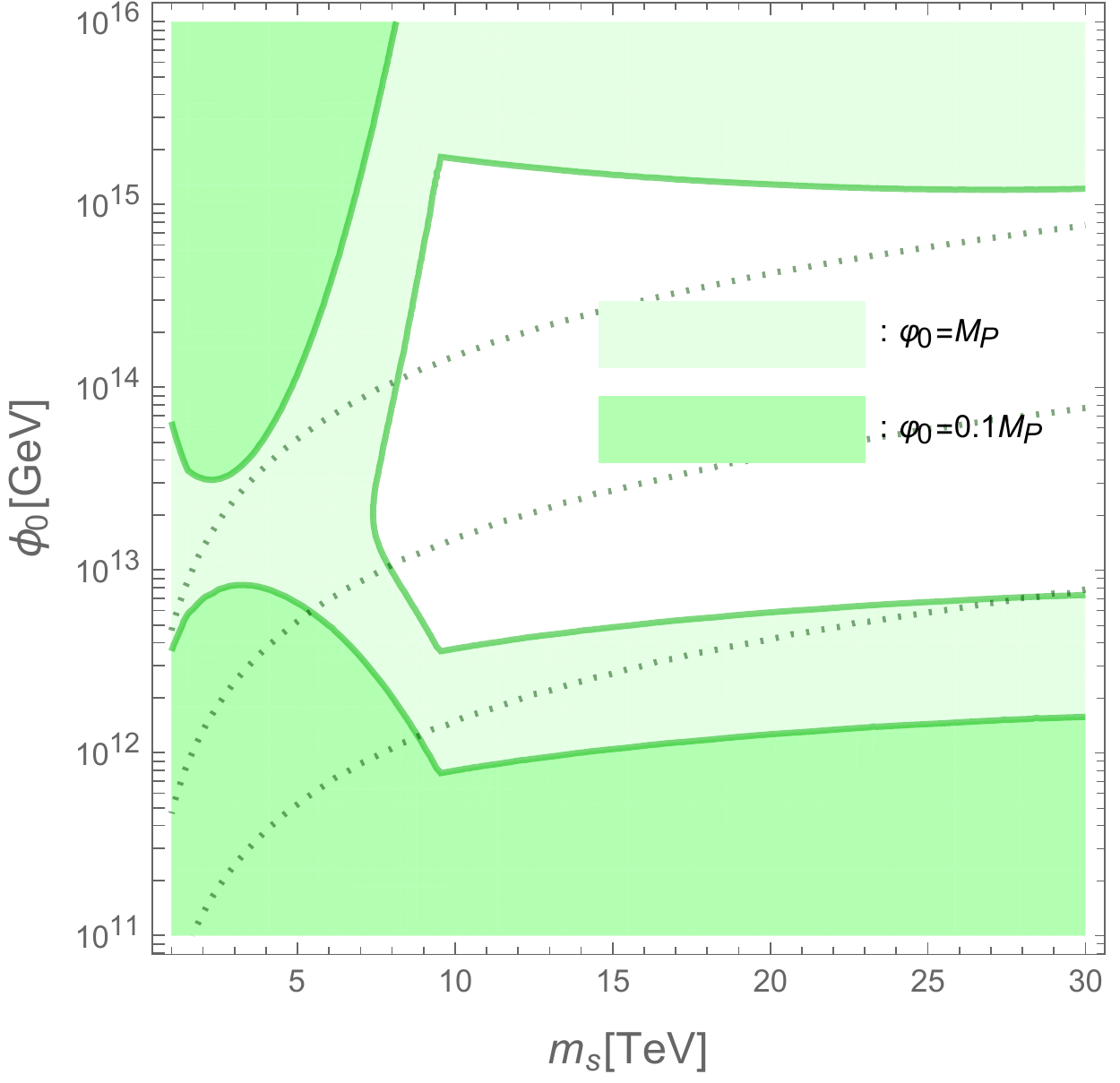}
\caption{
Parameter space of thermal inflation 
free from the cosmological moduli problem  (white region).
Dotted lines correspond to $T_{\rm d} = 1, 10, 100\, {\rm GeV}$ from top to bottom,  respectively. 
Here we have taken $m_\phi = |\mu| = m_s$, $m_A = 1.5 |\mu|$,  and 
$B = 2m_A/3$ as an example.
}
\label{fig:para-moduli}
\end{center}
\end{figure}

Fig.~\ref{fig:para-moduli} shows the parameter space of thermal inflation in which cosmological moduli problem is  resolved. 
One can see that for $m_\varphi \sim \mu \sim m_\phi \sim T_{\rm c} \sim m_s \gtrsim 1 {\rm TeV}$ as a natural possibility the symmetry-breaking scale of $U(1)_{B-L}$ is constrained to be in a range of
\beq \label{eq:phi0-range}
10^{12} {\rm GeV} \lesssim \phi_0 \lesssim 10^{15} {\rm GeV}
\eeq
Note that, if there were multiple thermal inflation processes \cite{Lyth:1995ka}, or if a single process
 is elongated by an extra source of radiation delaying the end of thermal inflation, the abundance of big-bang moduli can be suppressed by several orders of magnitude \cite{Choi:2012ye}.
In such a case, the symmetry-breaking scale $\phi_0$ can be much lower than the lower bound in \eq{eq:phi0-range} by a couple of orders of magnitude.
If $m_\varphi \gg \mu \sim m_\phi \sim T_{\rm c} \sim m_s$, the abundance of big-bang moduli can be reduced but only mildly since $Y_\varphi^{\rm BB} \propto m_\varphi^{_1/2}$, but the BBN constraint becomes much weaker, allowing a lower
 value of $\phi_0$.
%

\section{Baryon number asymmetry}
\label{sec:BAU}

Thermal inflation erases pre-existing particles as the primordial inflation does.
In the present scenario, 
the amount of dilution required to resolve the moduli problem is huge, 
and hence baryon asymmetry should be reproduced after thermal inflation in order to be compatible with observations. 
This can be achieved by the late time Affleck-Dine leptogenesis~\cite{Stewart:1996ai,Jeong:2004hy,Kim:2008yu,Park:2010qd}.
The processes discussed in Ref.~\cite{Park:2010qd} works in our scenario too in the same way except that the role of PQ field is played 
by the $B-L$ Higgses.  

The MSSM $\mu$-parameter in our scenario comes mainly from the higher order term
involving the Higgs bilinear  in \eq{eq:W-high}, implying  $\mu(\phi) \approx \lambda_\mu \phi^2/M$.
With the effective mass-squared parameter along the $LH_u$ flat direction defined as
\beq
m_{LH_u}^2(\phi) \equiv \frac{1}{2} \l( m_L^2 + m_{H_u}^2 + |\mu(\phi)|^2 \r),
\eeq
the late time Affleck-Dine mechanism  requires 
\beq
m_{LH_u}^2(\phi= 0) < 0 \quad{\rm and}
\quad 
m_{LH_u}^2(\phi=\phi_0) > 0,
\eeq
with
\beq
\l| m_\phi^2 \r| < \l| m_{LH_u}^2(\phi=0) \r|.
\eeq
Then, 
the $LH_u$ flat direction is destabilized before $U(1)_{B-L}$ is broken 
by $\phi$.
The symmetry is broken as $\phi$ is destabilized from the origin at $T_{\rm c} \sim \sqrt{\overline{m^2}}$, ending thermal inflation. 
As $\phi$ goes far away from the origin, satisfying $M_N = y_N \langle \phi \rangle \gg m_s$ but $m_{LH_u}^2(\phi) < 0$, the VEV of the $LH_u$ direction increases, 
 and is given by
\beq \label{eq:phi-AD}
\phi_{\rm AD} (\phi) \sim v_u \l( \frac{|m_{LH_u}^2(\phi)|}{A_\nu m_\nu} \r)^{1/2} \l( \frac{\phi}{\phi_0} \r)^{1/2},
\eeq
where $v_u$ is the VEV of the up-type Higgs, 
$m_\nu \lesssim \mathcal{O}(10^{-2}) {\rm eV}$ is the scale of the neutrino mass parameter in 
the flavor basis, and $A_\nu$ is the soft SUSY breaking $A$-parameter associated with 
the Weinberg operator appearing for $y_N |\phi| \gg m_s$.
Once $\phi$ reaches close to the vacuum,   the effective mass-squared of the $LH_u$ direction becomes positive, 
and as a result,  the flat direction is lifted up.
In the presence of  a $CP$ violating phase,  a motion along the phase direction of the complex field associated with the $LH_u$ flat direction is generated at the same time. 

The lepton number asymmetry produced right after the $U(1)_{B-L}$ phase transition 
is expected to be~\cite{Park:2010qd}
\beq \label{eq:YL-at-Tc}
Y_L^{\rm c} \sim \frac{\theta_{\rm CP} m_{\rm AD} \phi_{\rm AD, osc}^2}{T_{\rm c}^3},
\eeq
where $\theta_{\rm CP} $ is the $CP$ violation phase of
the $LH_u$ flat direction,  which we take as a free parameter, 
$m_{\rm AD} \equiv \sqrt{m_{LH_u}^2(\phi_0)}$, 
and $\phi_{\rm AD, osc} \sim \phi_{\rm AD}(\phi_0)$ is the oscillation amplitude of 
the $LH_u$ field at the onset of its oscillation soon after thermal inflation.
Note that $\phi_{\rm AD, osc}$ is approximately the largest one among those of three lepton flavors.
The late time asymmetry is then expected to be 
\beq
Y_L \sim \frac{\theta_{\rm CP}  T_{\rm d} m_{\rm AD} \phi_{\rm AD, osc}^2}{V_{\rm TI}} = \frac{\theta_{\rm CP}}{\beta^2} \frac{T_{\rm d}}{m_\phi} \frac{m_{\rm AD}}{m_\phi} \l( \frac{\phi_{\rm AD, osc}}{\phi_0} \r)^2,
\eeq
and the baryon asymmetry at the present universe reads
$Y_B = (10/31) Y_L$ from a given lepton flavor~\cite{Davidson:2008bu}.
Note that the LSP's are also produced in the eventual decay of the AD particle.
If the decay takes place well below the freeze-out temperature of the lightest MSSM superparticle, the mass of the dark matter is constrained to be around $\mathcal{O}(10)\, {\rm GeV}$.
However, this does not happen in our scenario due to  partial reheating right after thermal inflation.

\begin{figure}[h]
\begin{center}
\includegraphics[width=0.45\linewidth]{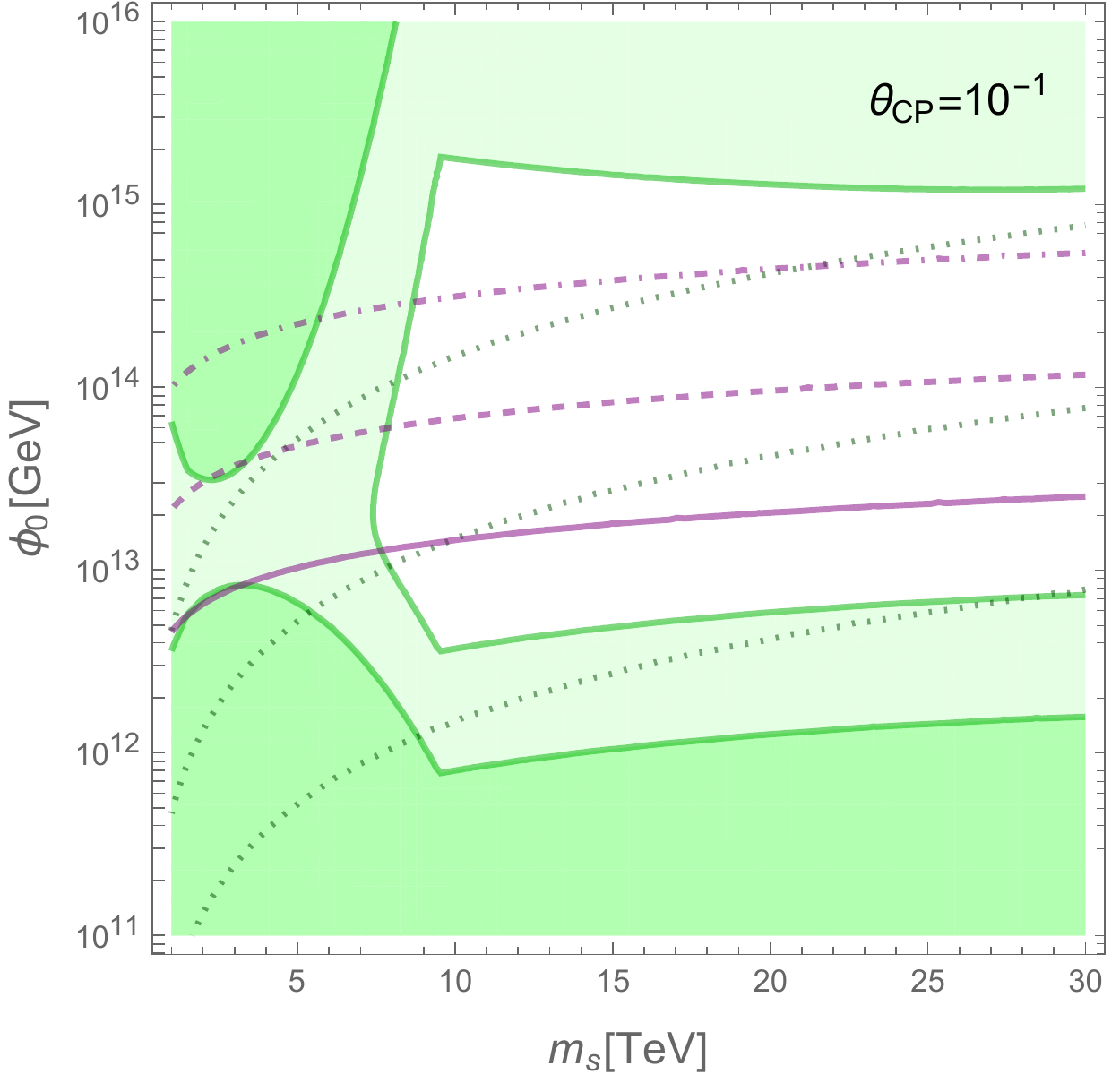}
\hspace{7pt}
\includegraphics[width=0.45\linewidth]{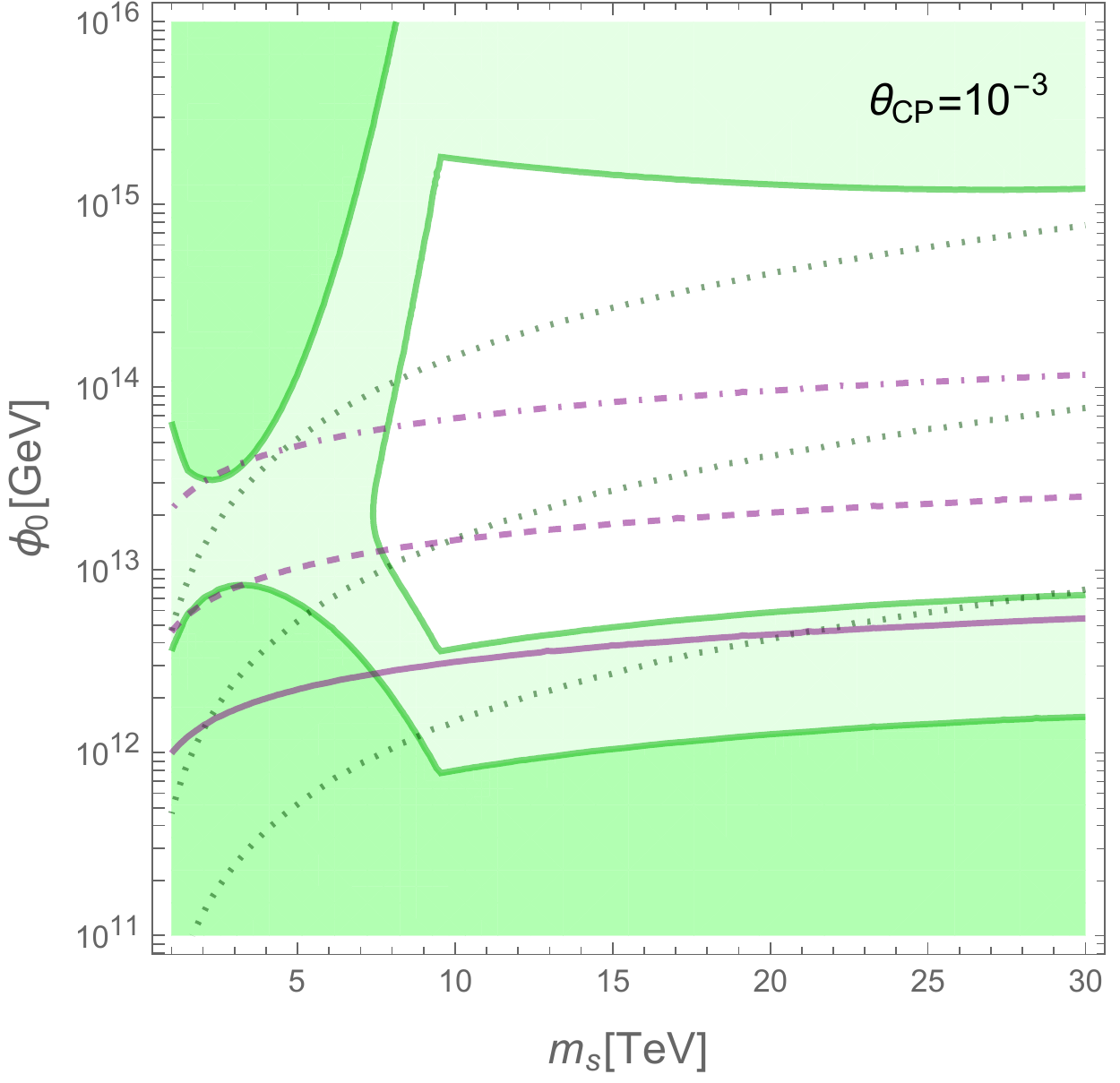}

\end{center}
\caption{
Parameter space for the right amount of baryon asymmetry,  where 
we have taken $\beta=1$, $m_{\rm AD}=A_\nu=m_s$,  and $\theta_{\rm CP} =10^{-1}$ (left), $10^{-3}$ (right).
Purple lines correspond  to $\phi_{\rm AD} = 10^{10}, 10^{11}, 10^{12} \,{\rm GeV}$ 
(or equivalently,  $m_\nu=3 \times 10^{-4}$, $3\times 10^{-6}$, $3\times 10^{-8}\, {\rm eV}$) from bottom to top.
The other color and line scheme are the same as in Fig.~\ref{fig:para-moduli}.
}
 \label{fig:para-BAU}
\end{figure}
In Fig.~\ref{fig:para-BAU},  we depict  the parameter space leading to
the right amount of baryon  asymmetry 
for several different values of $\phi_{\rm AD, osc}$.
As shown in the figure,  depending on the $CP$ violating phase and the mass parameter of neutrino flavor, 
the allowed value of $\phi_0$ can vary within a couple of orders of magnitude.

\section{Dark matter} 
\label{sec:DM}

There are various candidates for dark matter in our scenario: the lighter of the ordinary lightest sparticle and axino, or the axions.
To implement the axion solution to the strong CP problem, one can consider a simple KSVZ-type model where the PQ symmetry is spontaneously broken by a single PQ scalar.
If the PQ scalar is radiatively stabilized,  its fermionic partner,  the axino, obtains a loop suppressed mass compared to other supersymmetric particles, and thus becomes the LSP.

\subsection{Ordinary neutralino LSP}

For $H_{\rm c} \gg \Gamma_\phi$
with $H_{\rm c}$ being the Hubble parameter at $T=T_{\rm c}$
and $\Gamma_\phi$ being the total decay width of the flaton,
the contribution to the radiation density from flaton decays soon after the end of thermal inflation is given by
\beq \label{eq:radiation-from-flaton}
\rho_{\rm r} \simeq \frac{2}{5} \frac{\Gamma_\phi}{H} \rho_\phi \sim \l( \frac{M_{\rm P}}{\phi_0} \r) \rho_{\rm r, c},
\eeq
where $\rho_{\rm r, c}$ is the radiation density at the end of thermal inflation.
Hence, for $\phi_0$ at an intermediate scale, the temperature after thermal inflation can be higher than the critical temperature by a couple of orders of magnitude.
This means that,   even in the case with $m_\phi  < 2 m_{\tilde \chi}$ with $m_{\tilde \chi}$ being the LSP mass, it is possible to reproduce LSP's from thermal bath due to high enough background temperature after thermal inflation. 
If the decay temperature of the flaton is sufficiently low, the final abundance of LSP's is fixed when the LSP freezes out from thermal bath during the matter-domination era, and then diluted due to the later entropy release \footnote{If moduli and gravitinos are heavy enough, the late-time decay of those particles can result in a right amount of neutralino dark matter without thermal inflation \cite{Kohri:2005ru}.}.

The abundance of neutralino LSP's in terms of the yield at freeze-out is given by
\beq
Y_{\tilde \chi}^{\rm fo} = \frac{1}{4 \sqrt{2 \alpha_{\rm d}} x_{\rm fo}} \l( \frac{m_{\tilde \chi}}{M_{\rm P}} \r) \frac{1}{\langle \sigma v_{\rm rel} \rangle T_{\rm d}^2},
\eeq
where $x_{\rm fo} \equiv m_{\tilde \chi}/T_{\rm fo}$ with $T_{\rm fo}$ being the freeze-out temperature, and $\langle \sigma v_{\rm rel} \rangle$ is the thermally averaged annihilation cross section.
As a good enough approximation, $x_{\rm fo}$ is obtained by equating the annihilation rate to the expansion rate as   
\beq
x_{\rm fo} = \frac{5}{2} \ln x_{\rm fo} + \ln \l[ \langle \sigma v_{\rm rel} \rangle T_{\rm d}^2 \l( \frac{M_{\rm P}}{m} \r) \r] + \ln \l[ \frac{\sqrt{2 \alpha_{\rm d}}}{\alpha(T)}  \frac{g_{\tilde \chi}}{\l( 2 \pi \r)^{3/2}} \r],
\eeq
where $g_{\tilde \chi}$ is the number of degrees of freedom of ${\tilde \chi}$.
For instance, it gives $x_{\rm fo} \approx 18.5$ for $m_{\tilde \chi} = 1 {\rm TeV}$ with $\langle \sigma v_{\rm rel} \rangle = 10^{-9} {\rm GeV}^{-2}$.
The thermal-averaged annihilation cross sections of various neutralinos LSP are collected at Appendix~\ref{sec:app-ann}.

The late time decay of flaton injects  entropy to thermal bath, and causes a dilution of the freeze-out yield of the LSPs by a factor of 
\beq
\Delta = \l( \frac{\alpha_{\rm fo}}{\alpha_{\rm d}} \r) \l( \frac{T_{\rm fo}}{T_{\rm d}} \r)^5.
\eeq
Hence, the present abundance of neutralino LSP dark matter is expected to be
\beq \label{eq:mchiYchi}
m_{\tilde \chi} Y_{\tilde \chi} = \frac{1}{4 \sqrt{2 \alpha_{\rm d}} x_{\rm fo}} \l( \frac{m_{\tilde \chi}}{M_{\rm P}} \r) \frac{m_{\tilde \chi}}{\langle \sigma v_{\rm rel} \rangle T_{\rm fo}^2} \l( \frac{\alpha_{\rm d}}{\alpha_{\rm fo}} \r) \l( \frac{T_{\rm d}}{T_{\rm fo}} \r)^3. 
\eeq
If the neutralino LSP is the main component of dark matter, observations require $m_{\tilde \chi} Y_{\tilde \chi} \approx 5 \times 10^{-10} {\rm GeV}$,  putting  a constraint on $T_{\rm d}$ as 
\bea \label{eq:Td-constraint-DM}
T_{\rm d} 
&=& m_{\tilde \chi} \l[ 4 \sqrt{2} \l( \frac{\alpha_{\rm fo}}{\alpha_{\rm d}^{1/2}} \r) x_{\rm fo}^{-4} \l( m_{\tilde \chi} Y_{\tilde \chi} \r)^{\rm obs} M_{\rm P} \langle \sigma v_{\rm rel} \rangle \r]^{1/3}
\nonumber \\
&\simeq& 52 {\rm GeV} \l( \frac{m_{\tilde \chi}}{1 {\rm TeV}} \r) \l( \frac{ \langle \sigma v_{\rm rel} \rangle }{10^{-9} {\rm GeV}^{-2}} \r)^{1/3} \l( \frac{20}{x_{\rm fo}} \r)^{4/3}.
\eea
Note that for a given type of the LSP, $T_{\rm d}$ is nearly fixed as a function of $m_{\tilde \chi} \sim m_s$. 
Also, solving the cosmological moduli problem lower-bounds the energy density of thermal inflation.
As a result, the scale of $U(1)_{B-L}$ breaking and the strength of the gauge coupling are nearly fixed 
\footnote{If a unification of gauge couplings is assumed,  $g_{BL} = \mathcal{O}(0.1-1)$ is expected.
Otherwise, it can be much smaller than unity as long as $\lambda_N = \mathcal{O}(1)$ in order to provide a sizable thermal mass to the $B-L$ Higgses.}.

\subsection{KSVZ axino LSP}

In the simplest KSVZ realization of the axion solution to the strong CP problem, axino obtains its mass dominantly via one-loop radiative corrections, and so the mass is generally smaller than $m_s$ by about two orders of magnitude~\cite{Moxhay:1984am,Goto:1991gq}.
Its couplings to MSSM particles are also generated radiatively: couplings to gluons and gluinos at the one-loop level, and 
couplings to quarks and squarks at the tow-loop level~\cite{Covi:2002vw}.
If the heavy KSVZ quarks carry an electric charge, axino couples to the $U(1)_Y$ gauge bosons and binos at the one-loop level~\cite{Covi:2001nw}. 
Having those interactions, axino LSPs can be produced by both the freeze-in mechanism through decays of thermal particles and the decays of frozen-out NLSPs.

For $T_{\rm d} = \mathcal{O}(10-100) {\rm GeV}$ with $m_{\tilde \chi} \gtrsim \mathcal{O}(1) {\rm TeV}$, the thermal freeze-in contribution is dominated by the decays of thermal NLSPs.
It depends on the axion decay constant $f_a$, but for $f_a \gtrsim \mathcal{O}(10^{10}) {\rm GeV}$ it is expected to be negligible~\cite{Covi:2001nw} or comparable to the contribution from decays of frozen-out NLSPs.
The abundance of axinos from frozen-out NLSPs is given by~\cite{Covi:1999ty}
\beq
\Omega_{\tilde a} = \frac{m_{\tilde a}}{m_{\tilde \chi}} \Omega_{\tilde \chi}^{\rm fo} = \mathcal{O}(10^{-2}) \Omega_{\tilde \chi}^{\rm fo},
\eeq
where $\Omega_{\tilde \chi}^{\rm fo}$ is the would-be fractional contribution of the NLSP obtained from the freeze-out process.
Note that, as long as the heavy KSVZ quarks carry an electric charge,  the NLSP neutralinos can decay well before the epoch of BBN,
 and the abundance of the frozen-out NLSP's is not constrained by BBN \footnote{If  heavy KSVZ quarks are electrically neutral, NLSP neutralinos are expected to decay during the BBN epoch unless the gluino mass is larger than at least a few tens of ${\rm TeV}$ or the axion decay constant is in the range of the hadronic axion window. 
In such a case,  the axino cannot be the main component of dark matter in the present universe. 
A sizable amount of axinos can be produced thermally in the hadronic axion window, but it requires a reheating temperature well below the ${\rm GeV}$ scale in order to avoid overproduction~\cite{Redino:2015mye}.}.

 \begin{figure}[h]
\begin{center}
\includegraphics[width=0.31\textwidth]{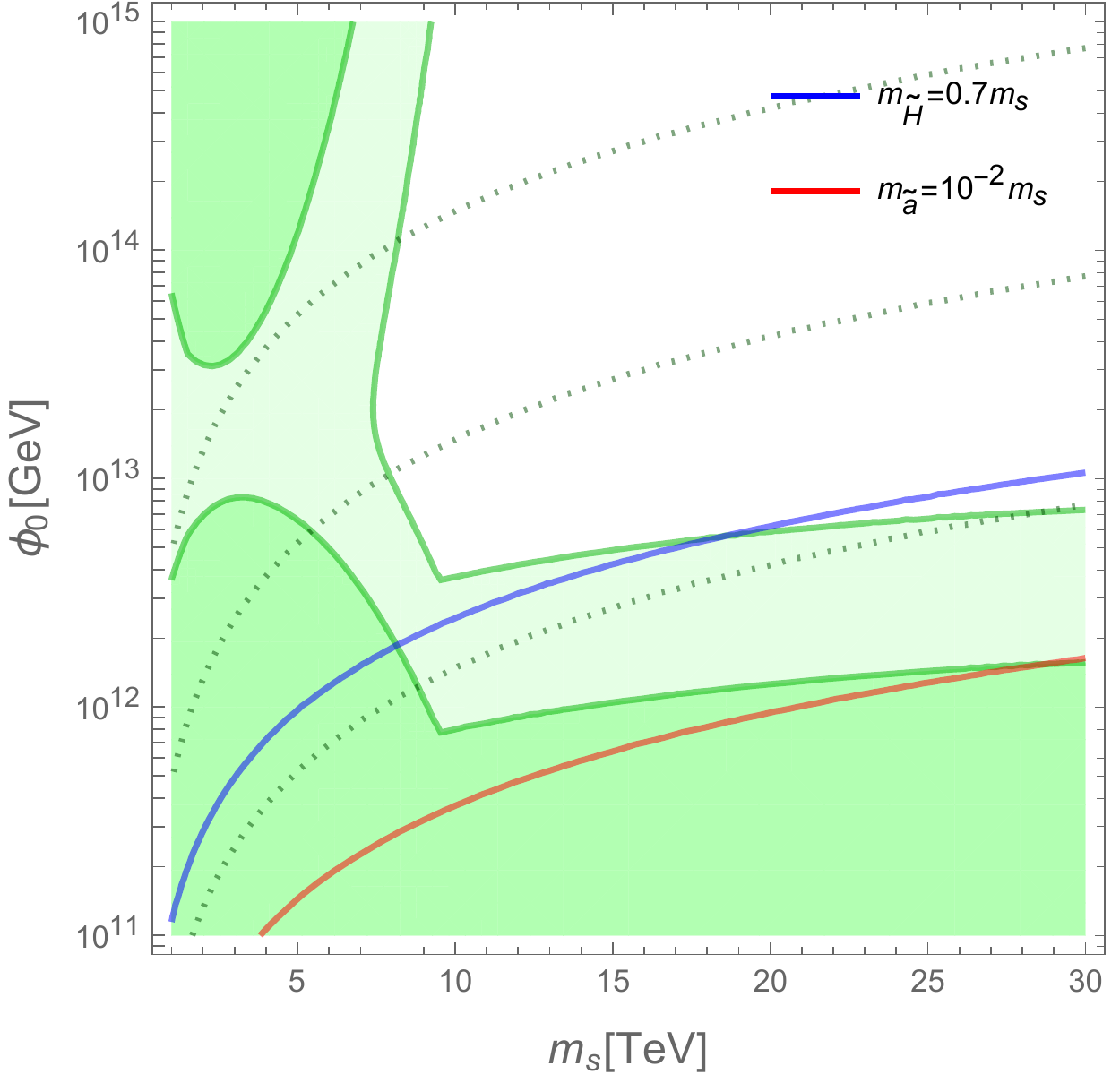}
\hspace{7pt}
\includegraphics[width=0.31\textwidth]{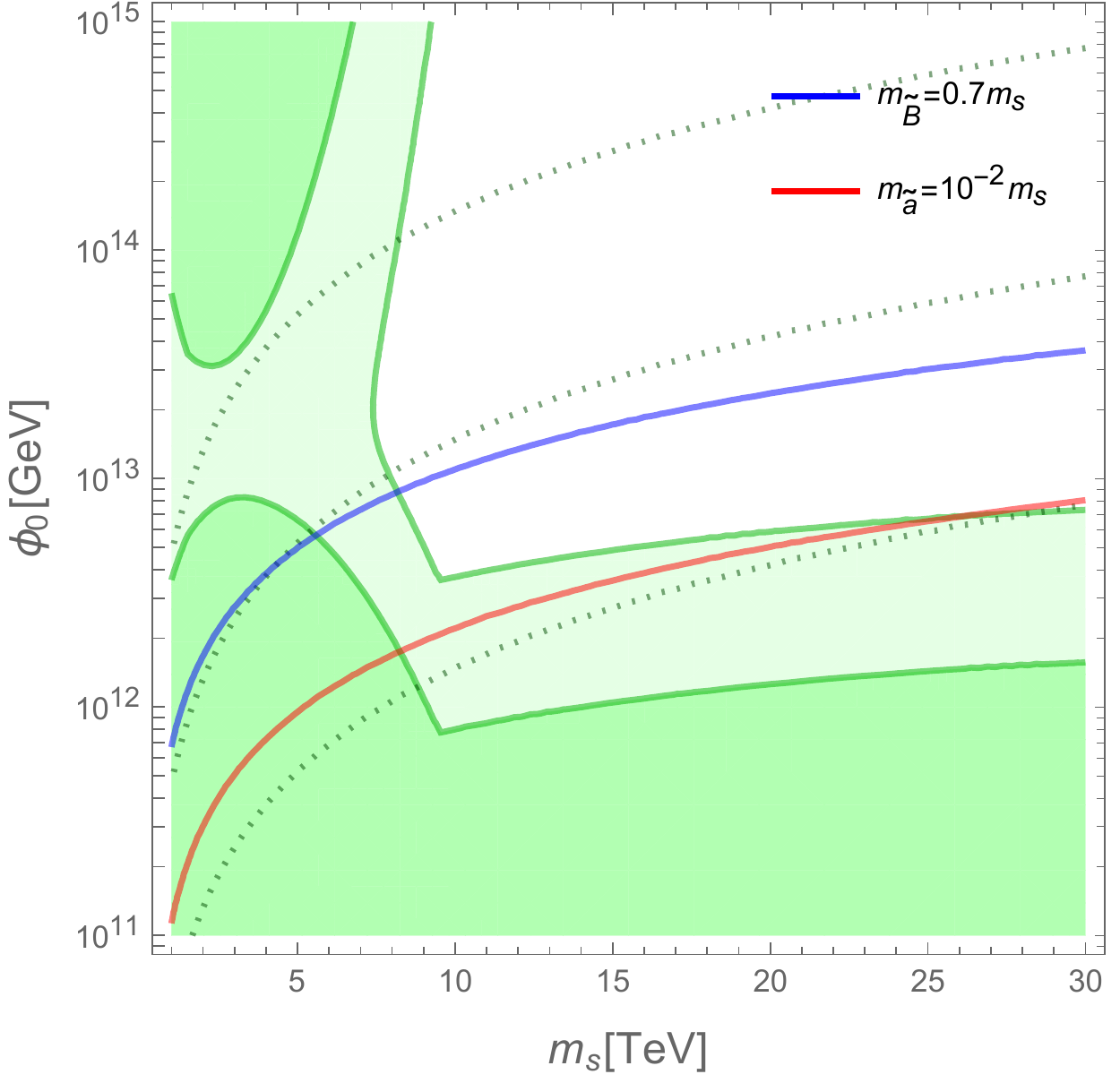}
\hspace{7pt}
\includegraphics[width=0.31\textwidth]{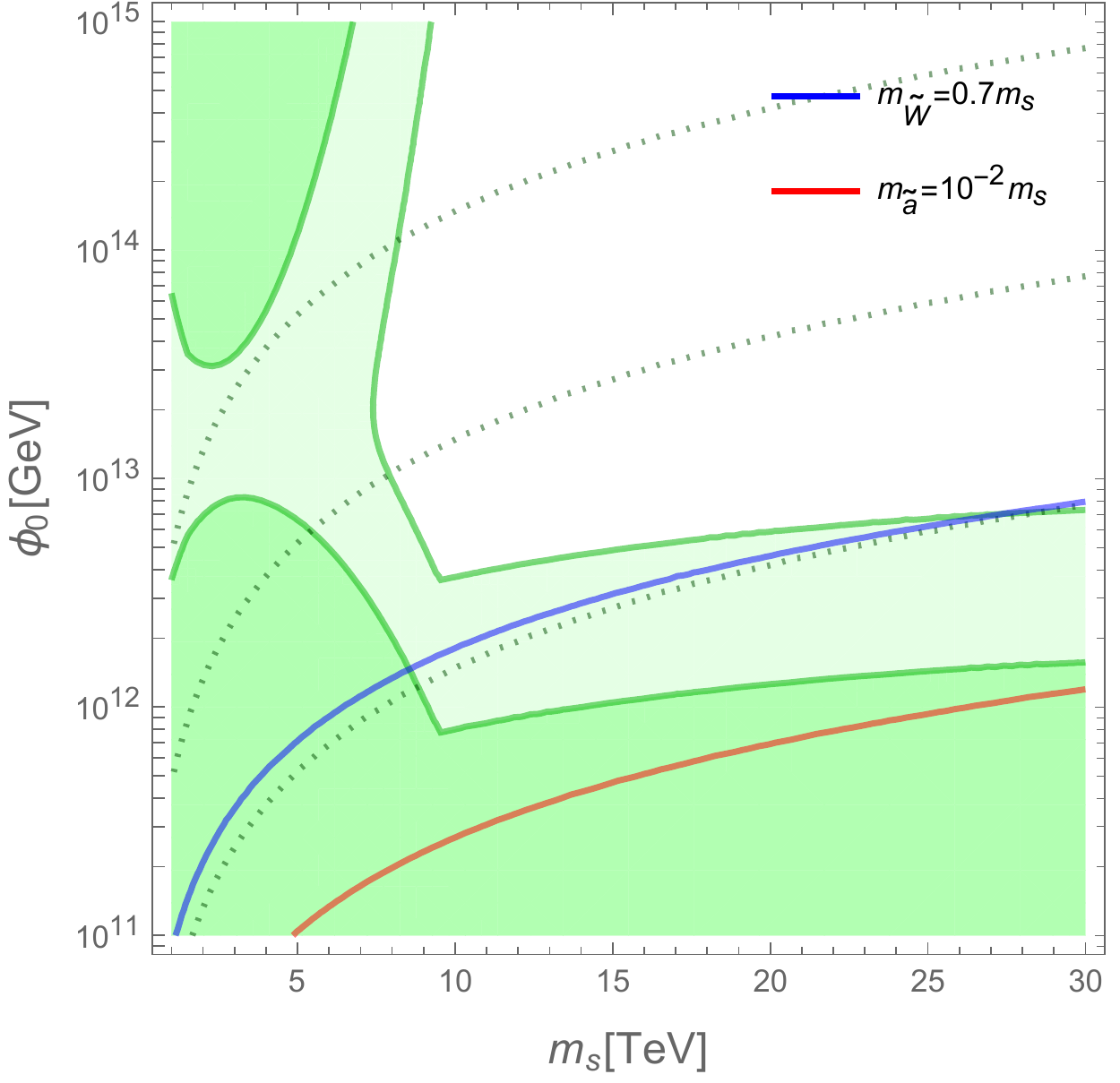}
\caption{
Parameter space for the right amount of dark matter with the same parameter set for 
$T_{\rm d}$ as in Fig.~\ref{fig:para-moduli}.
The panels from left to right correspond to the  Higgsino-LSP, bino-LSP, and wino-LSP  case, respectively.
In each panel, the blue and red lines are for the cases of neutralino LSP and axino LSP obtained from only frozen-out neutralino NLSPs as an example.
}
\label{fig:para-DM}
\end{center}
\end{figure}
In Fig.~\ref{fig:para-DM}, we show  the parameter spaces for the right amount of dark matter
from various types of the LSP. 
It is also possible to have the sneutrino LSP, and if it is mostly wino-like, the viable parameter space is expected to be in between those for
the cases of Higgsino and Bino LSP. 
As shown in \eq{eq:Td-constraint-DM}, the neutralino DM requires $T_{\rm d} = \mathcal{O}(10-100)\, {\rm GeV}$, which is obtained for $\phi_0 = \mathcal{O}(10^{12-14})\,{\rm GeV}$.
For axino LSPs,  $\phi_0$ can be lowered down by about an order of magnitude.
Note that, for $m_s = \mathcal{O}(10) {\rm TeV}$, axinos from Bino NLSPs are likely to be the only viable possibility.

\section{Stochastic gravitational wave backgrounds}

In our scenario, the local $U(1)_{B-L}$ symmetry is broken at the end of thermal inflation after primordial inflation.
The $B-L$ phase transition is expected to be of the first order type due to the flatness of the potential with order unity couplings of the flaton to thermal bath.
Also, the completion of the phase transition causes the formation of a network of local cosmic strings \cite{Vilenkin:2000jqa}.
Hence,  stochastic gravitational wave backgrounds can be generated from the first order phase transition through, 
for instance,   bubble collisions, sound waves, turbulences, and cosmic strings \cite{Caprini:2018mtu}~\footnote{As a thorough study of GW's from cosmic strings, see Ref.~\cite{Gouttenoire:2019kij} for example.}.

For an intermediate symmetry-breaking scale,   which is of our interest, 
the contribution to the energy density of gravitational waves from bubble collisions and accompanying phenomena is
 smaller than that from cosmic strings by many orders of magnitude.
This is because the former is a temporal event arising at the epoch of the phase transition, 
and suffers from a strong dilution due to a somewhat long period of the matter-domination era after thermal inflation 
till the eventual reheating due to the decay of flaton \cite{Easther:2008sx}.
So, here we consider the contribution from cosmic strings only.

There have been several works on GW's from cosmic strings formed from the $U(1)_{B-L}$ breaking.
However, most of them have considered strings whose core width is similar to the inverse of the symmetry breaking 
scale~\cite{Buchmuller:2013lra,Buchmuller:2019gfy,Chakrabortty:2020otp}.
Contrary to those cases, 
the strings appearing in our scenario have a 
much larger 
core width of the scalar field configuration 
than that of the gauge field configuration,
by many orders of magnitude \cite{Barreiro:1996dx,Perkins:1998re,Cui:2007js}~\footnote{See also Refs.~\cite{Hill:1987qx,Freese:1995vp,Penin:1996si,Donaire:2005qm} for other works on strings from flat directions.}.
The critical difference between those two types has been discussed in detail in  Ref.~\cite{Cui:2007js}.

It is known that, soon after its formation, a string network follows the so-called scaling solution keeping the fractional energy 
contribution of the network constant during the evolution of the universe \cite{Kibble:1984hp,Bennett:1987vf,Bennett:1989ak,Allen:1990tv}.
Such a behavior is due to the formation of sting loops \cite{Shellard:1987bv,Eto:2006db} and their energy loss via emissions of gravitational waves and energetic particles \cite{Vilenkin:2000jqa,Olum:1998ag,Matsunami:2019fss}.
Dominance of either the former or the latter in the energy loss depends on the length of string loops relative to 
the characteristic scales associated with the small-scale structure relevant for particle emissions, cusps and kinks \cite{Blanco-Pillado:1998tyu,Blanco-Pillado:2015ana}.

In Ref.~\cite{Blanco-Pillado:1998tyu}, considering (thin) cosmic strings of the conventional Abelian-Higgs model, 
authors have found that the scale of a cusp,  i.e., ~the over-lapping length,  is given by 
\beq
\ell_{\rm cusp} \approx \sqrt{w_s \ell},
\eeq
where $w_s$ and $\ell$ are the width (i.e., the core radius) and the length of a string loop, respectively.
The power of particle emissions from a cusp of a loop oscillating with relativistic speed is then found as
\beq \label{eq:P-cusp}
P_{\rm cusp} \sim N_c \mu_s \l( \frac{w_s}{\ell} \r)^{1/2},
\eeq
where $N_c \sim 2$ is the number of cusps in a string loop \cite{Blanco-Pillado:2015ana}, and $\mu_s$ is the string tension, 
i.e.~the energy per unit length.
In the case of kinks, particle emissions are from kink-kink collisions.
The over-lapping region of two kinks is about $w_s^3$.
The emission power is thus given by~\cite{Matsunami:2019fss}
\beq \label{eq:P-kink}
P_{\rm kink} \sim N_k \mu_s \l( \frac{w_s}{\ell} \r) \sim \frac{N_k}{N_c} \l( \frac{w_s}{\ell} \r)^{1/2} P_{\rm cusp},
\eeq
where $N_k$ is the average number of kink-kink collisions and might be as large as $\mathcal{O}(10^3)$~\cite{Ringeval:2017eww}.
Note that, whereas $w_s$ is of a micro-physical scale, $\ell $ is of the cosmological scale, 
and in our scenario $P_{\rm cusp} \gg P_{\rm kink}$.
So, we ignore the effect of kinks in the discussion from now on. 
For gravitational waves, the emission power reads~\cite{Vilenkin:2000jqa}
\beq \label{eq:P-GW}
P_{\rm GW} = \Gamma G \mu_s^2,
\eeq
where $\Gamma \sim 50$ is the emission coefficient of gravitational waves, and $G$ is the Newton's constant.
From the comparison to the emission power of gravitational waves, the characteristic length of a string loop for the domination of particle emissions is found as \cite{Auclair:2019jip}
\beq \label{eq:l-star}
\ell_* = \beta_{\rm c} w_s / \l( \Gamma G\mu_s \r)^2,
\eeq
where $\beta_{\rm c}$ is a numerical constant of order unity.

Compared to the case of thin strings from the conventional Abelian-Higgs model, cosmic strings from a supersymmetric flat direction have several critical differences \cite{Barreiro:1996dx,Perkins:1998re,Cui:2007js}.
First of all, the core radii of associated gauge and scalar fields are quite different to each other.
Denoted as $r_V$ and $r_\phi$ respectively, the radii are given by $r_V \sim m_V^{-1}$ and $r_\phi \sim m_\phi^{-1}$.
Since $m_V \sim g_{BL} \phi_0$ with $g_{BL}$ being the associated gauge coupling and $m_\phi \sim m_s$, one finds 
\beq
r_V \lll r_\phi \ \l({\mbox{ i.e.}} \  m_V \ggg m_\phi \r),
\eeq
unless $g_{BL}$ is smaller than unity by many orders of magnitude. 
The scalar field configuration of this type of cosmic strings has a long log-tail in the region $r_V \ll r_s \lesssim r_\phi$ with $r_s$ being the distance from the center of the string core \cite{Cui:2007js}.
Secondly, because of the fact $r_V \lll r_\phi$,   a pair of strings feel an attractive force 
irrespectively  of the orientation of strings,   when they get close to each other.
Under kinetic constraints due to energy conservation, such a force allows to form a string segment of higher winding numbers 
as long as its tension is lower then the sum of the two strings involved.
Since it is energetically favored,  once formed, such a segment can grow like a zippering.  
This zippering effect results in equally distributed string species characterized by their winding numbers upto $N_w^{\rm max}(t)$ which is numerically found as \cite{Cui:2007js}
\beq \label{eq:NwMax}
N_w^{\rm max}(t) \approx N_c \l( \frac{t}{t_c} \r)^{0.22},
\eeq
where $N_c$ is the typical winding number of strings formed at the phase transition ending thermal inflation at $t_c$, and $t$ is a later comic time.
The net magnetic flux trapped by surrounding bubbles may be responsible for $N_c$, leading to \cite{Rajantie:2001ps,Blanco-Pillado:2007ihs}
\beq
N_c \sim \frac{g_{BL}}{2 \pi} \sqrt{R_b T_c} \longrightarrow \frac{g_{BL}}{2 \pi} \sqrt{\frac{\beta M_{\rm P}}{\phi_0}} \sim \mathcal{O}(1),
\eeq
where $R_b$ is the typical size of bubbles at their coalescence, and in the far right-hand side of the equation we have used $R_b H_c \equiv \beta = \mathcal{O}(10^{-3})$
as  expected in the case of the phase transition along a flat direction having order unity couplings to thermal bath \cite{Easther:2008sx}, 
and $H_c \sim m_\phi \phi_0/M_{\rm P}$ with $T_{\rm c} \sim m_\phi$.
For the range of parameters of our interest, the maximal winding number reaches to $N_w^{\rm max}(t_0) \lesssim \mathcal{O}(10^8)$.
The zippering increases or decreases the winding number $N_w$ of the zipped segment, and results in a dependence of the string tension on $N_w$ as \cite{Cui:2007js}
\beq \label{eq:mu-Nw}
\frac{\mu_s(N_w)}{\pi \phi_0^2} 
\approx c_1 \l( 1 + c_2 \ln N_w \r), 
\eeq
with the coefficients roughly given by 
\bea \label{eq:coeff-c1}
c_1 &\approx& \frac{4.2}{\ln \l( \frac{m_V^2}{m_\phi^2} \r)} + \frac{14}{\ln^2 \l( \frac{m_V^2}{m_\phi^2} \r)}, 
\\ \label{eq:coeff-c2}
c_2 &\approx& \frac{2.6}{\ln \l( \frac{m_V^2}{m_\phi^2} \r)} + \frac{57}{\ln^2 \l( \frac{m_V^2}{m_\phi^2} \r)},
\eea
and $c_1 \sim c_2 \sim 0.1$ for $m_\phi \sim m_s= \mathcal{O}(1)\, {\rm TeV}$ and $\phi_0 = \mathcal{O}(10^{12-13})\, {\rm GeV}$ with $g_{BL} = \mathcal{O}(0.1)$ as an example.
Consequently, 
it leads to the $N_w$ dependence of the emission power $P_{\rm GW}$ in \eq{eq:P-GW}.
We however  ignore  the zippering effect on loops once formed (see for example Ref.~\cite{Firouzjahi:2009nt}).

For a given species of string equilibrated with a specific $N_w$, one can find the relation of the present time frequency $f$ of GW's to the background temperature $T$ of the universe when the loops contributing dominantly to the signal is formed, as follows.
Numerical simulations show that, when they are formed at a cosmic time $t_i$ from a network of strings, the typical length of loops which
is 
the most relevant for GW-emissions is $\xi t_i$ with $\xi= \mathcal{O}(0.1)$ \cite{Blanco-Pillado:2013qja}.
The length of such string loops at later time is then given by
\beq
\ell(t) = \xi t_i - \Gamma G \mu_s \l( t - t_i \r).
\eeq
Also, it has been noticed  that the loop formulation rate has a behavior of $dn /dt_i \propto t_i^{-4}$ with $n$ being the number density of a type of loops \cite{Gouttenoire:2019kij}.
This implies that the dominant contribution to GW's at a frequency $f$ at the present universe is from the associated loops formed at the earliest possible epoch.
Based on this observation, 
one finds that,  
if a loop is formed during the standard radiation-domination era after the reheating of thermal inflation, the temperature-to-frequency relation reads \cite{Gouttenoire:2019kij}
\beq \label{eq:T-to-f}
f \simeq \sqrt{\frac{8}{z_{\rm eq} \xi \Gamma G \mu_s}} \l( \frac{g_*(T)}{g_*(T_0)} \r)^{1/4} \l( \frac{T}{T_0} \r) t_0^{-1},
\eeq
where $z_{\rm eq}$ is the red-shift at the matter-radiation equality, $g_*(T)$ is the number of relativistic degrees of freedom at $T$, $t_0$ is the age of the universe, and the recent dark energy domination was ignored.

Using \eq{eq:T-to-f}, the length scale $\ell_*$ given in \eq{eq:l-star} for a string species with a given winding number can be related to the associated background temperature $T_*$ as follows.
Taking $\ell_* = \xi t_*$ with the expansion rate given by \eq{eq:H-phi-dom}, we find that string loops of size $\ell_*$ start appearing after the reheating if
\beq \label{eq:TstarCut-cond}
m_\phi > 4 \l( \frac{\pi \xi}{3 \beta_c}  \r)^{1/2} \l| 1 - \frac{|B|^2}{m_A^2} \r|^{-1} \l( \frac{m_\phi}{\mu} \r)^2 \frac{\Gamma G \mu_s \phi_0}{\sqrt{w_s m_\phi}}.
\eeq
Comparing to the parameter space for the right amount of dark matter and baryon  asymmetry discussed in the previous sections, 
we notice that,  in most of the relevant parameter space, 
$\ell_*$ appears after the flaton decay,  i.e.~during the standard radiation-domination epoch recovered after thermal inflation.
Paying attention to this case only, we find that the corresponding background temperature $T_*$ at the first formation of string loops with the size $\ell_*$ is given by
\beq \label{eq:T-star}
T_* 
= \l( \frac{\xi}{2 \alpha_*^{1/2} \beta_{\rm c} w_s m_\phi} \r)^{1/2} \l( \Gamma G \mu_s \r) \l( m_\phi M_{\rm P} \r)^{1/2},
\eeq
where $\alpha_* = \alpha(T_*)$.
Note that this is different from the case of the conventional Abelian-Higgs model, since the string width $w_s$ of the flat-direction strings has nearly nothing to do with the string tension $\mu_s$.
From \eq{eq:T-to-f}, the associated characteristic frequency, $f_*$, from which a change of the spectrum of the gravitational waves starts to appear, 
can be found as \cite{Gouttenoire:2019kij}
\beq \label{eq:f-star}
f_* \simeq 6.7 \times 10^{-2} {\rm Hz} \l( \frac{T_*}{{\rm GeV}} \r) \l( \frac{0.1 \times 50 \times 10^{-11}}{\xi \Gamma G \mu_s} \r)^{1/2} \l( \frac{g_*(T_*)}{g_*(T_0)} \r)^{1/4}.
\eeq
Note that $T_* \propto \mu_s$ leads to $f_* \propto \mu_s^{1/2}$.
That is, $f_*$ depends on $N_w$.
But loops of a given size $\ell$ are populated mostly at its earliest possible epoch.
Hence, $N_w$-dependence of $f_*$ is at most from $N_w^{\rm max}(t_*)$ with $t_*$ being the cosmic time when loops of $\ell_*$ starts forming.
Using \eqs{eq:NwMax}{eq:mu-Nw} with $t_*/t_{\rm c} \sim \l( M_{\rm P}/\phi_0 \r)^3$, we find that $f_*$ varies at most about $40$\%.
For simplicity,   we ignore this variation of $f_*$ in the estimation of GWs without affecting the main point of our argument.

Meanwhile, if $T_{\rm d} < T_*$ which is the case \eq{eq:TstarCut-cond} is not satisfied, a careful analysis shows that, taking into account of the time-dependence of loop-formation efficiency which is affected by changes of cosmology, a fractional deviation relative to the one expected in the standard cosmology starts appearing at \cite{Gouttenoire:2019kij}
\beq \label{eq:fd}
f_{\rm d} 
\simeq {\rm Hz} \l( \frac{T_{\rm d}}{{\rm GeV}} \r) \l( \frac{0.1 \times 50 \times 10^{-14}}{\alpha \Gamma G \mu_s} \r)^{1/2} \l( \frac{g_*(T_{\rm d})}{g_*(T_0)} \r)^{1/4}
\times \l\{
\begin{array}{ll}
6.3 \times 10^{-2} & \textrm{for VOS, 10\%}
\\
1.3 \times 10^{-3} & \textrm{for VOS, 1\%}
\end{array}
\r.
\eeq
where ``VOS'' stands for the ``Velocity-dependent One-Scale'' equations describing the evolution of a network of long strings.
Note that, if the sensitivity of a detector is good enough, even for $T_* < T_{\rm d}$, $f_{\rm d}$ can be smaller than $f_*$.

The fractional energy density of SGWB's is defined as 
\beq
\Omega_{\rm GW}(f) \equiv \frac{f}{\rho_{\rm c}} \frac{d \rho_{\rm GW}(f, t_0)}{df},
\eeq
where $\rho_{\rm c}$ is the critical energy density at the present universe, and $\rho_{\rm GW}$ is the energy density of the gravitational waves at a frequency $f$ at the present time $t_0$.
In addition to the contributions from string loops, there can be bursts of gravitational waves from individual cusps.
However, for $G \mu \lesssim \mathcal{O}(10^{-10})$,   the former is more promising \cite{Damour:2000wa,Damour:2001bk,Siemens:2006vk,Hogan:2006we,Siemens:2006yp}, and we consider only the contribution.
In Ref.~\cite{Gouttenoire:2019kij}, authors have provided a useful formula for the estimation of $\Omega_{\rm GW}$. 
It is given as
\beq \label{eq:GW-master-eq1}
\Omega_{\rm GW}(f) = \sum_k \Omega_{\rm GW}^{(k)}(f),
\eeq
with
\beq \label{eq:GW-master-eq2}
\Omega_{\rm GW}^{(k)}(f) 
= 
\frac{1}{\rho_{\rm c}} \frac{2 k}{f} \frac{\mathcal{F}_\xi \Gamma^{(k)} G \mu_s^2}{\xi \l( \xi + \Gamma G \mu_s \r)}
\int_{t_{\rm osc}}^{t_0} d{\tilde t} \frac{C_{\rm eff}(t_i)}{t_i^4} \l[ \frac{a({\tilde t})}{a_0} \r]^5 \l[ \frac{a_i}{a({\tilde t})} \r]^3 
\Theta(t_i - t_{\rm osc}) \Theta ( t_i - \ell_*/\xi),
\eeq
where $t_{\rm osc}$ is the starting time of long string oscillations, 
and $t_i$ the loop formation time given by
\beq \label{eq:t-formation}
t_i(f, {\tilde t}) = \frac{1}{\xi + \Gamma G \mu_s} \l[ \frac{2k}{f} \frac{a({\tilde t})}{a_0} + \Gamma G\mu_s {\tilde t} \r],
\eeq
with ${\tilde t}$ being the time at the emission of GWs whose frequency is $f$ at the present universe.
We refer the reader to Ref.~\cite{Gouttenoire:2019kij} for the details of the \eqs{eq:GW-master-eq1}{eq:GW-master-eq2}.

Contrary to the Abelian-Higgs model, \eqs{eq:GW-master-eq1}{eq:GW-master-eq2} are not directly applicable to the flat-direction cosmic strings due to the zippering effects, i.e., the $N_w$ dependence. 
Each string species has different lifetime and emission power of GWs.
Since the dependence is not simply factorized, strictly speaking, in order to take the zippering effects into account, we have to sum up all the contributions from various string species up to $N_w^{\rm max}(t)$ at a given cosmic time $t$, and integrate over time. 
Instead of such direct summing up, 
we note that the change in $\mu_s$ is at most less than about a factor 3 relative to the case of $N_w=1$ and $f \propto \mu_s^{-1/2}$ (see \eq{eq:T-to-f}), 
and take an approximation in which the $N_w$ dependence is ignored in $f$ but encoded only in the emission power averaged over string species of different $N_w$s.
Although it is not precise, this approach can show at least qualitatively the impact of the zippering effects on the spectrum of GWs.
The averaged power is given by 
\beq \label{eq:P-average}
\overline{P}_{\rm GW}(t) = \frac{1}{N_w^{\rm max}(t)} \sum_{N_w = 1}^{N_w^{\rm max}} P_{\rm GW}(\mu_s(N_w))
\xrightarrow[]{N_w^{\rm max} \ggg 1} \Gamma G \mu_s^2(N_w^{\rm max}(t)),
\eeq
which we apply at the time string loops start forming ($t_i$).
Equipped with this, we replace \eq{eq:GW-master-eq2} with
\bea \label{eq:GW-master-modified-approx}
\overline{\Omega_{\rm GW}^{(k)}}(f) 
 &\equiv& 
\frac{1}{\rho_{\rm c}} \frac{2 k}{f} \frac{\mathcal{F}_\xi \Gamma^{(k)} G \mu_{s,c}^2}{\xi \l( \xi + \Gamma G \mu_{s,c} \r)} \int_{t_{\rm osc}}^{t_0} d{\tilde t} \l( 1 + c_2 \ln N_w^{\rm max}(t_i) \r)^2
\nonumber \\
&\times& \frac{C_{\rm eff}(t_i)}{t_i^4} \l[ \frac{a({\tilde t})}{a_0} \r]^5 \l[ \frac{a_i}{a({\tilde t})} \r]^3 
\Theta(t_i - t_{\rm osc}) \Theta ( t_i - \ell_*/\xi),
\eea
where $\mu_{s,c} \equiv \mu_s(N_w=1)$, and it should be understood that the $N_w$ dependence is only 
with $N_w^{\rm max}(t_i)$ in the integrand.
Since the emission becomes stronger at the later universe, the lower the frequency is, the more enhanced the signal is.
Such enhancements make a qualitative difference of the GW-spectrum in comparison to the case without zippering effects.

The most stringent current bound on the fractional energy density of SGWBs from cosmic strings is from PPTA data \cite{Blanco-Pillado:2017rnf},
\beq
\Omega_{\rm GW} h^2(f_{\rm PPTA}) < 10^{-10} ,
\eeq 
where $f_{\rm PPTA}=2.8 \times 10^{-9} {\rm Hz}$.
This bound corresponds to
\beq
G \mu_s < 1.5 \times 10^{-11}.
\eeq
for Nambu-Goto type strings if the string intercommutation probability is unity. 
In our scenario,  due to the enhancement caused by zippering effects,  we can translate the bound as
\beq \label{eq:phi0-Up-bnd}
\phi_0 < \sqrt{\frac{\mu_{s, \rm GT}^{\rm bnd}}{\pi c_1 \zeta^{1/2}}},
\eeq
where $\mu_{s, \rm GT}^{\rm bnd}$ is the bound on the string tension for Nambu-Goto type strings, and the $\phi_0$-dependent enhancement factor $\zeta(\phi_0)$ is defined as
\beq
\zeta \equiv \frac{\Omega_{\rm GW}(f_{\rm PPTA})}{\Omega_{\rm GW, c}(f_{\rm PPTA})},
\eeq
with $\Omega_{\rm GW, c}$ being the expected GW's in the absence of zippering effects.

%
\begin{figure}[th] 
\begin{center}
\includegraphics[width=0.495\linewidth]{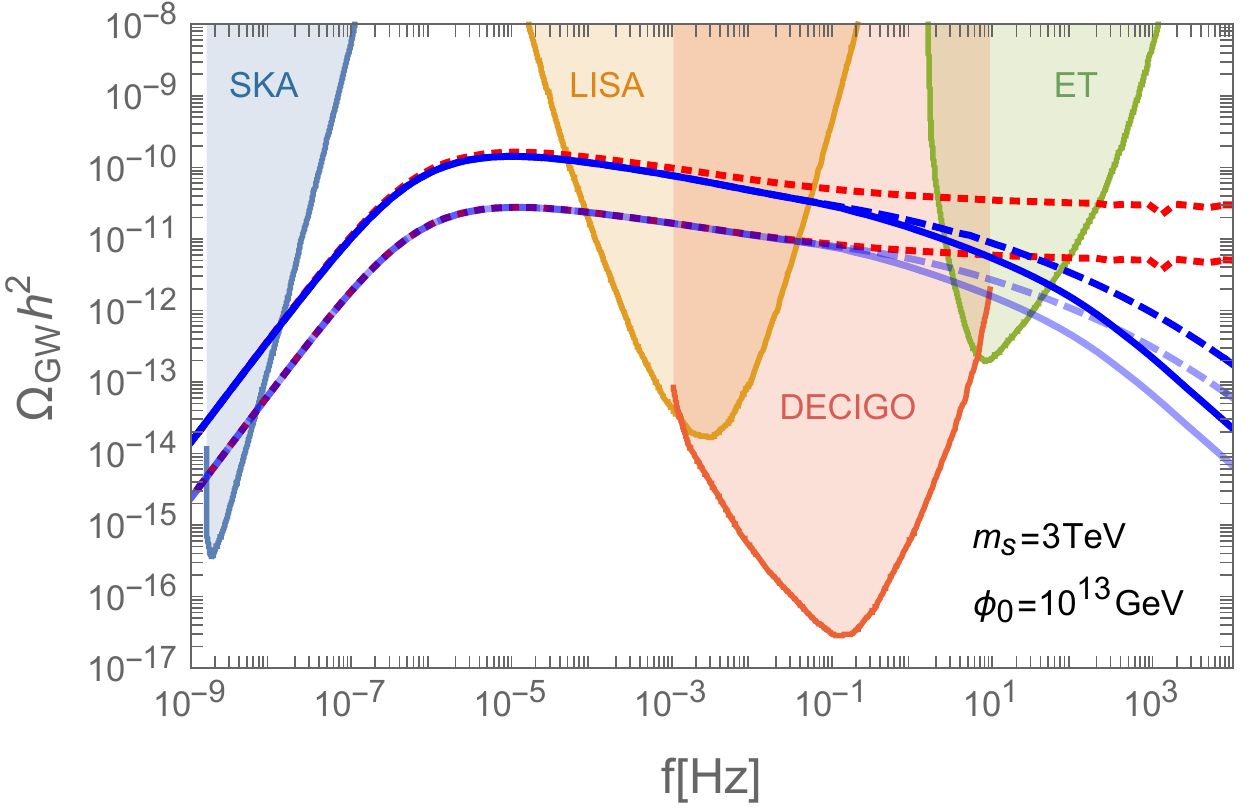}
\hspace{10pt}
\includegraphics[width=0.465\linewidth]{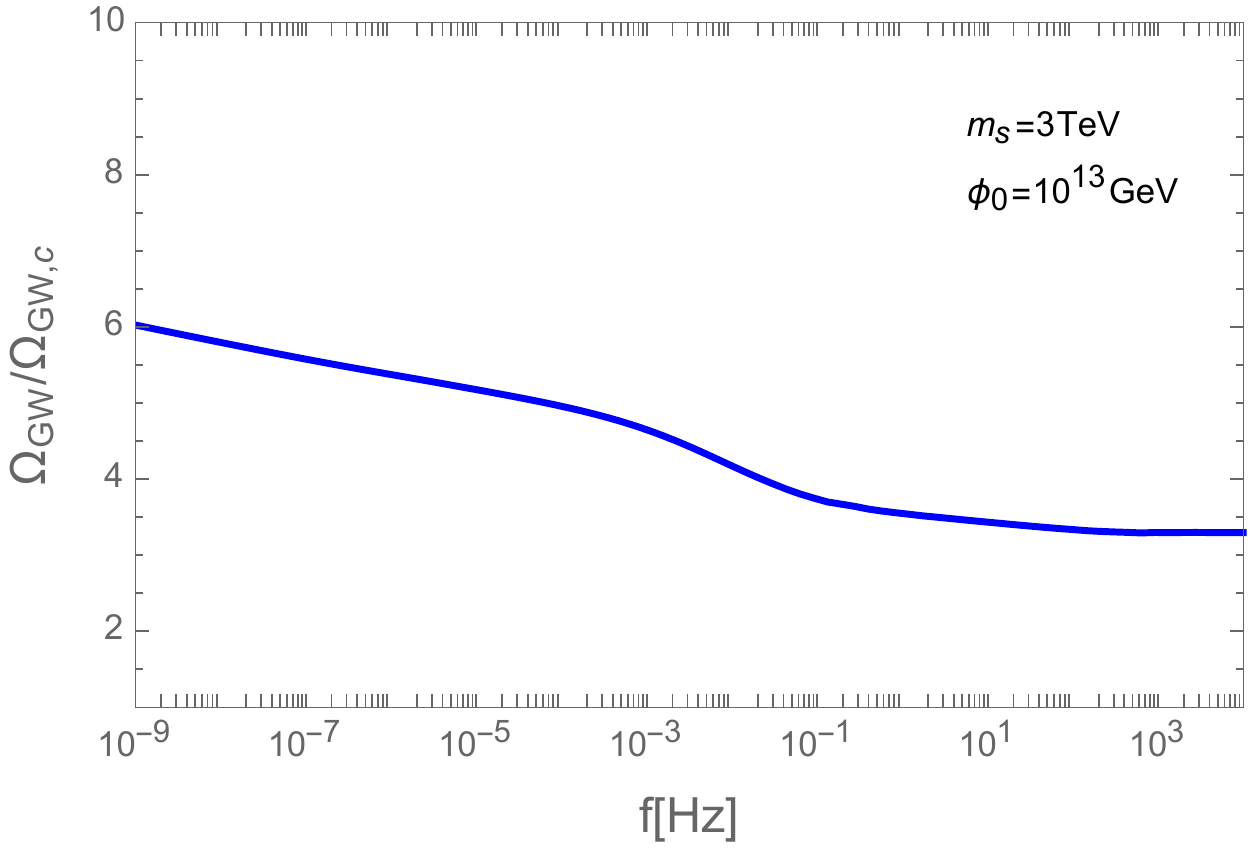}
\end{center}
\caption{
\textit{Left}: Forecasts of SGWB's for $m_\phi = m_s = 3 {\rm TeV}$ and $\phi_0 = 10^{13} {\rm GeV}$ corresponding to $G \mu_{s,c} \simeq 2.3 \times 10^{-13}$ .
The blue lines correspond to the case of thermal inflation without and with zippering effects from bottom to top, respectively.
The solid and dashed lines are with and without the small scale effect (i.e., the cutoff due to the cusp scale $\ell_*$).
The lower red dotted line is the case of the conventional Abelian-Higgs model without zippering effects in the standard cosmology.
The upper red dotted line is the same as the lower red dotted line but normalized to be matched exactly to the blue solid line at $f=f_{\rm PPTA}$ just for the comparison to see a spectral deviation. 
\textit{Right}: The ratio of signals shown in the left panel (blue to light-blue line in the case of thermal inflation with the small scale effect).
}
\label{fig:GWs-w-wo-zipper-ms3T}
\end{figure}
In Fig.~\ref{fig:GWs-w-wo-zipper-ms3T}, we depict  the expected spectrums of SGWB's in the presence of thermal inflation with and without zippering effects,
 and the case of Abelian-Higgs model without zippering in the standard cosmology (left panel) \footnote{In the case of the conventional Abelian-Higgs model, the cusp effect was ignored, since it appears at very high frequency region.}.
For the evolution of string network, we followed VOS model from Abelian-Higgs simulations with particle production \cite{Correia:2019bdl}.
From the figure,   we see that the zippering effects cause  a significant enhancement of the signal.
The enhancement is frequency-dependent as is clearly shown in the right panel of the figure.
Such a frequency-dependent enhancement may be probed at least LISA and DECIGO, and become a clear signal of the presence of a flat-direction.
Also, note that the turning point frequency associated with the characteristic scale of cusps or the reheating temperature after thermal inflation can be found in the end at the planned DECIGO experiment.
It would provide the information of the flaton mass which comes directly from soft SUSY breaking parameters.
Hence, in the sense that flat-directions with VEV's at intermediate scales arise naturally in SUSY theories, if detected, such SGWB's may be regarded as an indirect hint of SUSY.

%
\begin{figure}[th] 
\begin{center}
\includegraphics[width=0.495\linewidth]{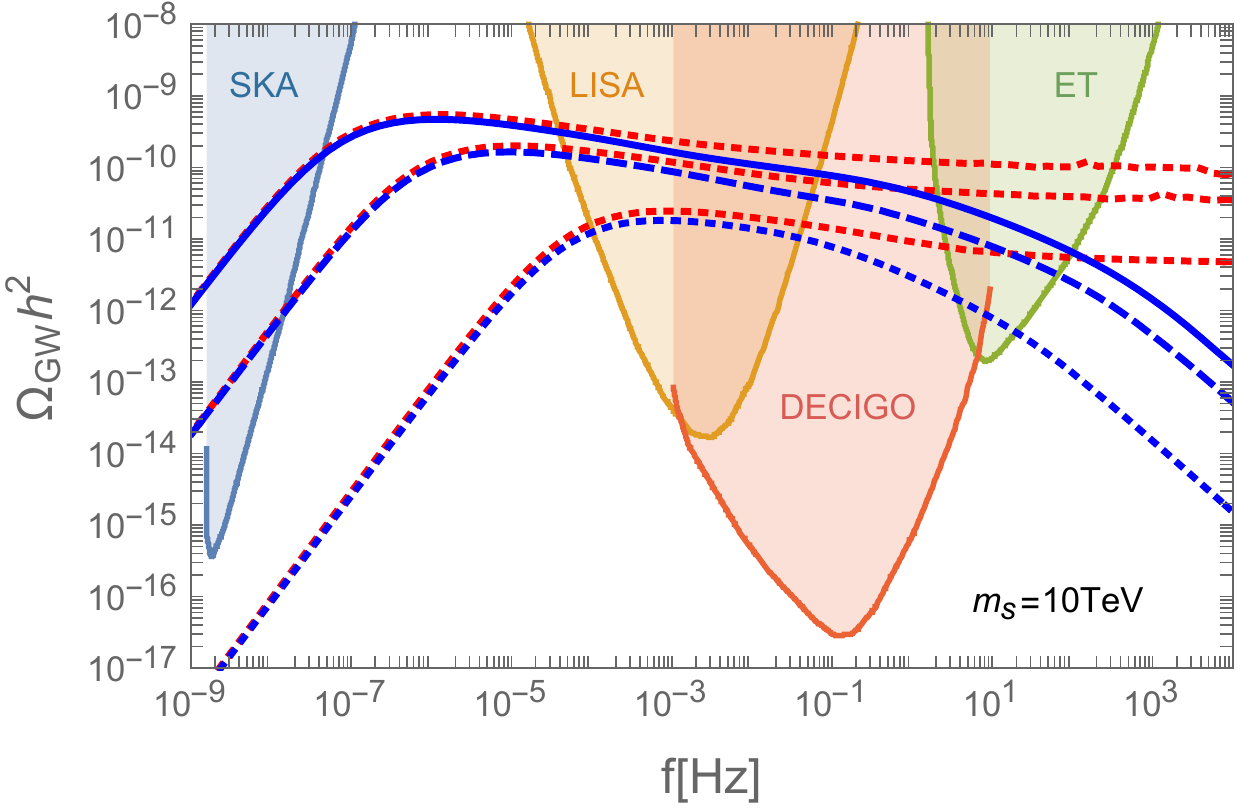}
\hspace{10pt}
\includegraphics[width=0.465\linewidth]{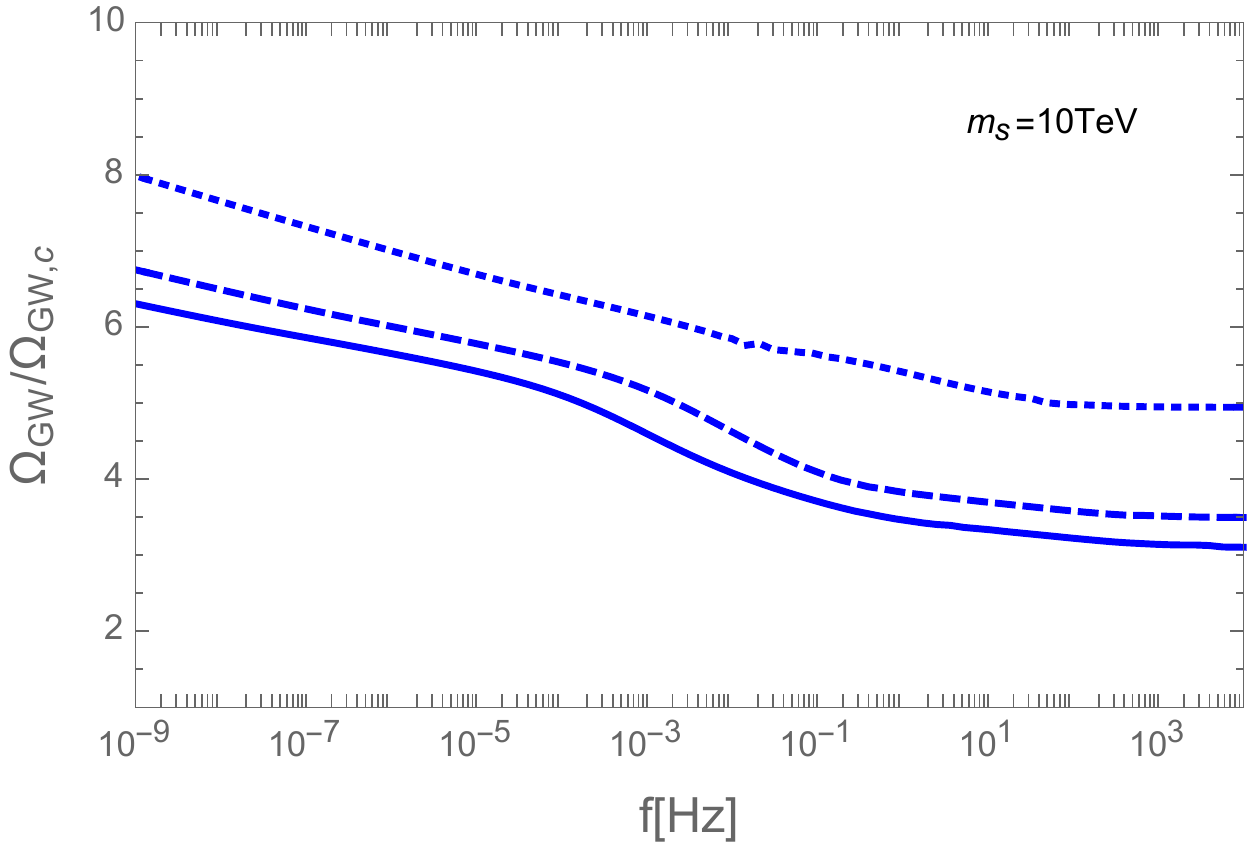}
\end{center}
\caption{
\textit{Left}: Forecasts of SGWBs for $m_\phi = m_s = 10\, {\rm TeV}$.
The blue lines correspond to the case of thermal inflation with zippering effect for $\phi_0 = 10^{12}\, {\rm GeV}, 10^{13} \,{\rm GeV}$, and $3 \times 10^{13} \,{\rm GeV}$ from bottom to top, respectively.
The red dotted lines are the case of the conventional Abelian-Higgs model without zippering effects in the standard cosmology but normalized to be matched exactly to each blue solid line at $f=f_{\rm PPTA}$ just for the comparison. 
\textit{Right}: The enhancement factor for each case of $\phi_0$ in the left-panel.
}
\label{fig:GWs-w-zipper-ms10T}
\end{figure}
In the left panel of Fig.~\ref{fig:GWs-w-zipper-ms10T}, we depicted expected signals of SGWBs for $m_\phi = m_s = 10 {\rm TeV}$ with $\phi_0 = 10^{12} {\rm GeV}, 10^{13} {\rm GeV}$ and $3 \times 10^{13} {\rm GeV}$ from bottom to top.
The largest $\phi_0$ nearly saturates the bound in \eq{eq:phi0-Up-bnd}.
As can be seen from the figure in all cases the signal of our scenario can be distinguished from the conventionally expected one.
In the right panel of the figure, the frequency-dependent enhancement in each case was depicted.
Note that, as $\phi_0$ increases, the enhancement effect decreases slightly.
This pattern can be understood from the $\phi_0$-dependence of $\mu_s$ in \eqs{eq:mu-Nw}{eq:coeff-c2} and that of $N_w^{\rm max}(t_i)$ with \eq{eq:T-to-f}.
Roughly speaking, for a given frequency an increase of $\phi_0$ causes a decrease of the coefficient $c_2$ in \eq{eq:coeff-c2}  while the change in $N_w^{\rm max}$ is minor. 
Hence, the enhancement effect becomes weaker.

\begin{figure}[th] 
\begin{center}
\includegraphics[width=0.48\linewidth]{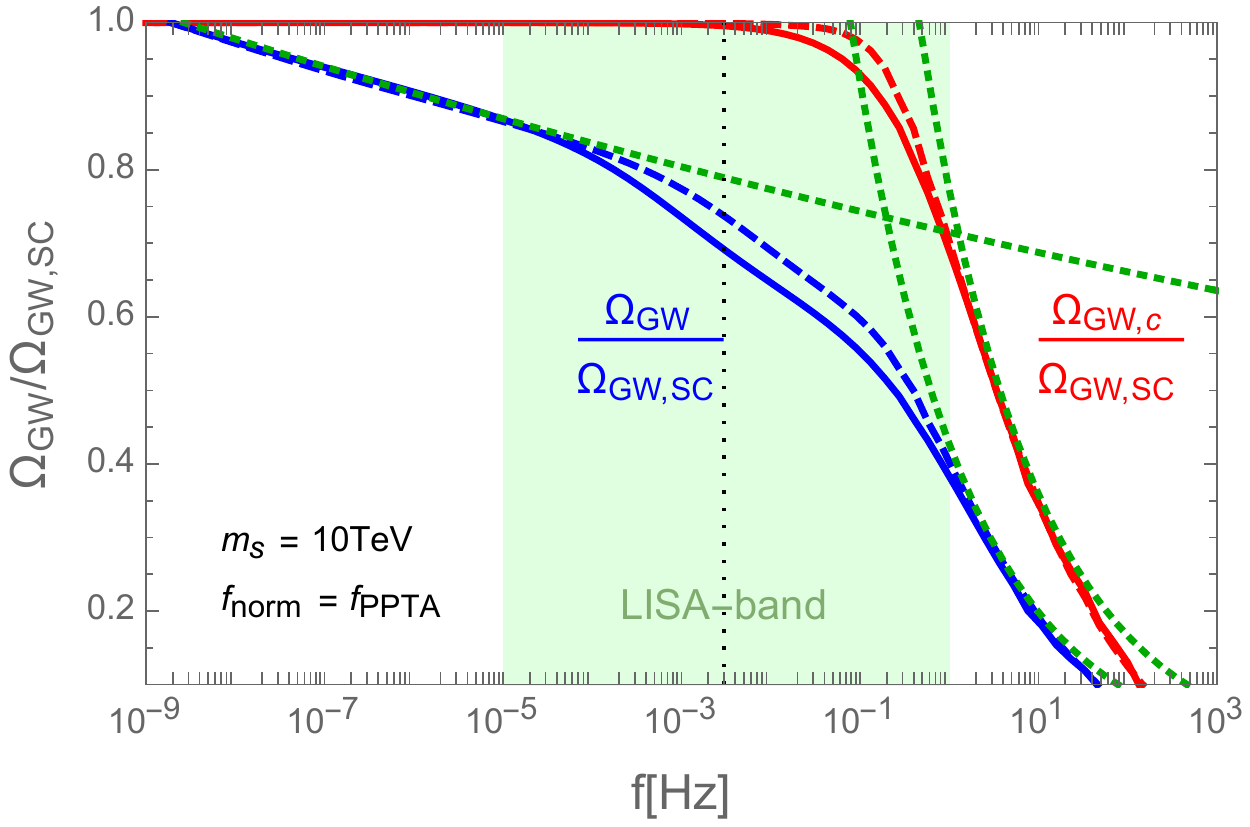}
\hspace{4pt}
\includegraphics[width=0.47\linewidth]{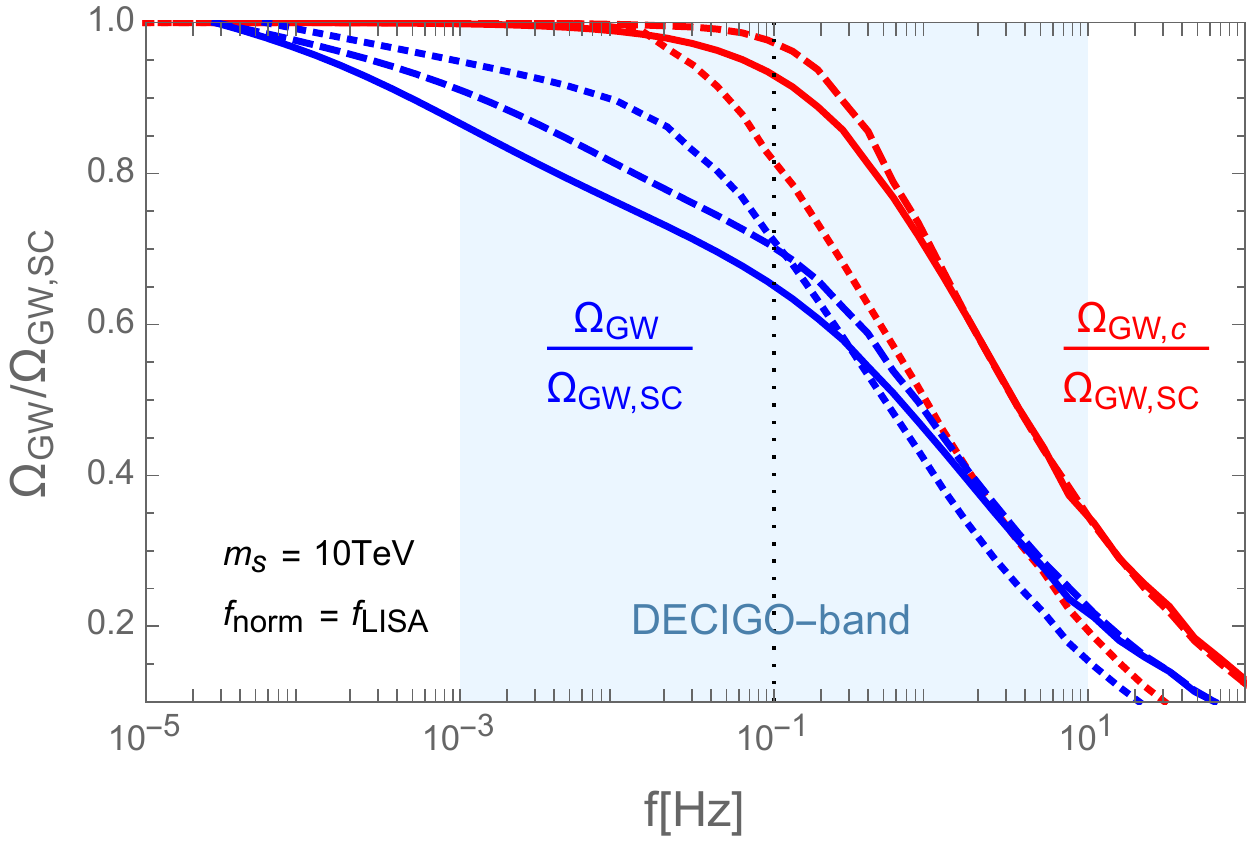}
\includegraphics[width=0.485\linewidth]{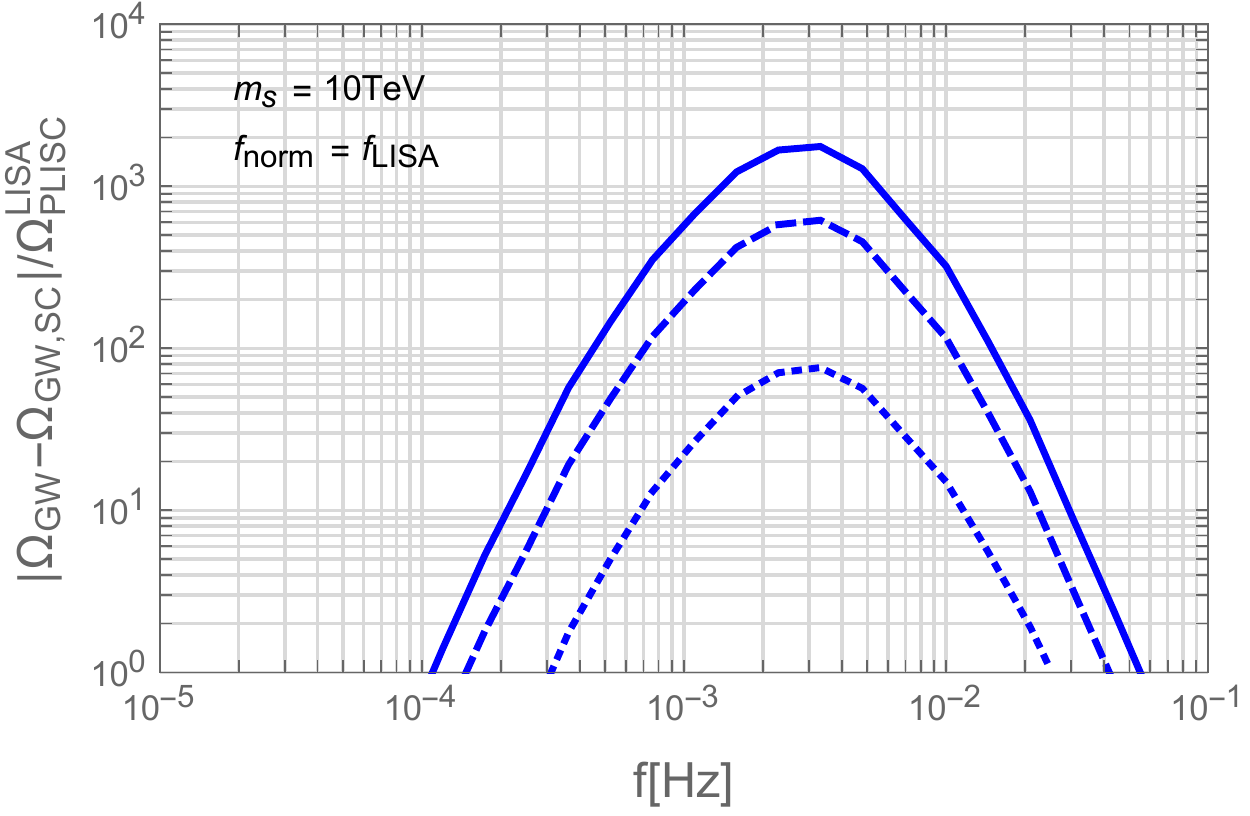}
\hspace{4pt}
\includegraphics[width=0.485\linewidth]{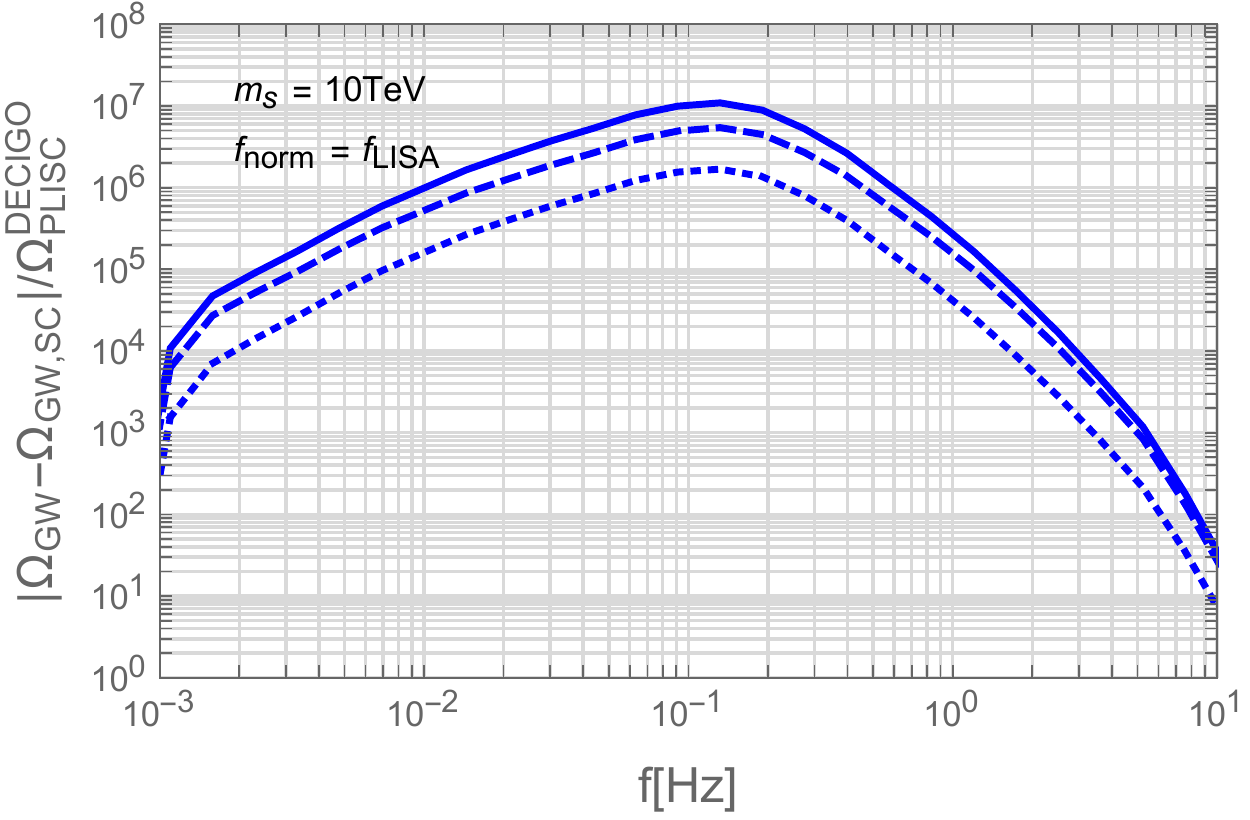}
\end{center}
\caption{
\textit{Top-Left}: Spectral deviations of signals in our scenario relative to the case in the standard cosmology without zippering effects ($\Omega_{\rm GW, SC}$) normalized at $f_{\rm PPTA}$ for comparisons for $m_s = 10\, {\rm TeV}$.
Red lines are without zippering effects ($\Omega_{\rm GW,c}$).
Blue lines are with zippering effects ($\Omega_{\rm GW}$).
Dashed and solid lines are for $\phi_0 = 10^{13} {\rm GeV}$ and $3 \times 10^{13} {\rm GeV}$, respectively.
The case for $\phi_0 = 10^{12} {\rm GeV}$ is not shown in the panel since the corresponding signal is out of the reach of the forthcoming PTA experiments such as SKA.
The dotted green lines are lines with specific power-law spectral indices matching the signals in specific bands (see descriptions in the text). 
The light-green shaded region is the LISA sensitivity band with the most sensitive frequency indicated as the vertical dotted line at $f \approx 3 \times 10^{-3} {\rm Hz}$. 
\textit{Top-Right}: The same as the top-left panel, but the case of $\phi_0 = 10^{12} {\rm GeV}$(dotted lines) was included and signals were normalized at $f_{\rm LISA}=10^{-4}{\rm Hz}$, $5 \times 10^{-5} {\rm Hz}$, and $4\times 10^{-5} {\rm Hz}$ for lines from bottom to top in each color.
$\Omega_{\rm GW}$ crosses $\Omega_{\rm PLISC}^{\rm LISA}$ curve at the normalization frequency of each case.
The light-blue shaded region is the DECIGO sensitivity band with the most sensitive frequency indicated as the vertical dotted line at $f \approx 10^{-1} {\rm Hz}$. 
\textit{Bottom-Left}: $\l| \Omega_{\rm GW} - \Omega_{\rm GW,SC} \r| / \Omega_{\rm PLISC}^{\rm LISA}$ with $\Omega_{\rm PLISC}^{\rm LISA}$ being the \textit{power-low-integrated-sensitivity-curve}(PLISC) of LISA for $t_{\rm obs} = 1{\rm yr}$ as the observing time with the \textit{SNR} threshold taken to be unity \cite{Schmitz:2020syl}.
It can be regarded as the signal-to-noise ratio of the deviation relative to the case of the standard cosmology without zippering effect at LISA experiment.
Line scheme and normalizations of signals are the same as the ones in the top-right panel.
\textit{Bottom-Right}: $\l| \Omega_{\rm GW} - \Omega_{\rm GW,SC} \r| / \Omega_{\rm PLISC}^{\rm DECIGO}$ at the DECIGO band.
Line scheme and normalizations of signals are the same as the ones in the bottom-left panel.
}
\label{fig:Probing-TI}
\end{figure}
The plausibility of probing our scenario in forthcoming experiments can be seen in Fig.~\ref{fig:Probing-TI}.
In the top-left panel of the figure, we depicted the ratio $\Omega_{\rm GW}/\Omega_{\rm GW,SC}$ to show the spectral deviation relative to the expectation of the standard cosmology for each different parameter set.
Signals were normalized at $f_{\rm PPTA}$ for clear comparisons.
It is applicable if signals are detected at forthcoming PTA experiments such as SKA.
From the red lines corresponding to the cases without zippering effects, one can clearly see the spectral bending caused by the cutoff due to small-scale structure (cusps) and/or by a change of cosmology (flaton-domination after thermal inflation to the standard radiation-domination).
Blue lines which include zippering effects show additional drastic changes compared to red lines even at frequencies below cutoff frequencies.
The non-trivial changes of spectrum around the central LISA sensitivity band seems due to an extra impact caused by the change of relativistic degrees of freedom, $g_*(T)$, encoded in $N_w^{\rm max}(t_i)$ in \eq{eq:GW-master-modified-approx}.
The green dotted lines are examples of the power-law fitting of the deviations.
We found that the deviation shows a spectral behavior of $f^{-\delta}$ with $\delta \approx 0.017$ at
the low-frequency region as the impact of zippering effects.
At the high-frequency region, 
it exhibits the behavior of  $f^{-1/3}$ 
as expected due to the cutoff of signals.
The deviation at high frequency end is due to the fact that high $k$-modes were summed upto $k = 2 \times 10^3$ 
just because of the limitation of efficiency in numerical calculations (see Ref.~\cite{Gouttenoire:2019kij} for a discussion of cutting out higher modes).  

In the top-right panel, $\Omega_{\rm GW}/\Omega_{\rm GW,SC}$ was depicted again, but each signal associated with a set of parameter was normalized at a frequency at which $\Omega_{\rm GW}$ crosses $\Omega_{\rm PLISC}^{\rm LISA}$, the \textit{power low integrated sensitivity curve}(PLISC) of the LISA experiment. 
This type of comparison would be relevant when only LISA data is available.
Note that the bending frequencies associated with high-frequency cutoff effects are all within the DECIGO band shown as the light-blue region, and not far away from the frequency of the peak sensitivity.

Probing the spectral deviations depends on the sensitivity of a detector.
In order to see the possibility of distinction of signals at LISA experiment which covers $10^{-5} \lesssim f/{\rm Hz} \lesssim 1$, in the bottom-left panel of the figure, we show a quantity, $\l| \Omega_{\rm GW} - \Omega_{\rm GW,SC} \r| / \Omega_{\rm PLISC}^{\rm LISA}$.
It can be regarded as the signal-to-noise ratio (\textit{SNR}) of the expected deviation relative to the one expected in the standard cosmology without zippering effects. 
One can see that the \textit{SNR} is much larger than unity in the frequency band $10^{-3} \lesssim f/{\rm Hz} \lesssim 10^{-2}$ for all different cases of $\phi_0$.
It means that it would be possible to discriminate the signal of thick strings experiencing the zippering effects.
In the bottom-right panel, we depicted $\l| \Omega_{\rm GW} - \Omega_{\rm GW,SC} \r| / \Omega_{\rm PLISC}^{\rm DECIGO}$ relevant for the DECIGO experiment.
Here 
we have used the same normalization of signals as the bottom-left panel, assuming LISA data would be already available when DECIGO data becomes available.
From the panel, we expect that, as long as it is located in the band $10^{-4} \lesssim f/{\rm Hz} \lesssim 1$, the bending frequency can be pinned down either at LISA or at least at the DECIGO experiment.

Distinction of the impact of small-scale structure from the low-reheating temperature of thermal inflation might be difficult, since they result in a same spectral pattern.
However, fitting a given data in the $(m_s, \phi_0)$-parameter space, one would be able to see which effect is likely to be responsible for the spectral bending of the GW signals.
Especially, since $f_* \propto m_\phi^{1/2} \phi_0$ and $f_{\rm d} \propto m_\phi^{3/2} / \phi_0^2$ as dominant parametric dependences, those two characteristic cutoff frequencies appear to be quite different to each other except some cases in which they are accidentally close, as shown in Table~\ref{tab:char-freq}.  
Note that, in all cases of $\phi_0$ shown in the table $f_{e\pm} < f_{\rm d}^{(1)} < f_{\rm QCD}$.
This means that, although $f_{\rm d}^{(1)}$s in some cases fall in the band where LISA experiment is sensitive enough to probe them, they become obscure due to the overwhelming effect caused by the change of $g_*(T)$ (see the top two panels in Fig.~\ref{fig:Probing-TI}).
Hence, practically, the lower of $f_*$ and $f_{\rm d}^{(10)}$ would appear as the detectable high-frequency cutoff if it is larger than $f_{\rm QCD}$.
\begin{table}[htp]
\begin{center}
\begin{tabular}{|c||c|c|c|}
\hline
\backslashbox{$f \l[ {\rm Hz} \r]$ }{$\phi_0$} & $10^{12} {\rm GeV}$ & $10^{13} {\rm GeV}$ & $3\times 10^{13} {\rm GeV}$ \\
\hline
\hline
$f_*$ & $0.025$ & $0.23$ & $0.67$ \\
\hline
$f_{\rm d}$ & $1.3\times 10^3$ & $14$ & $1.6$ \\
\hline
\multirow{2}{*}{$f_{\rm d}^{(10)}$} & $38$ & $0.40$ & $0.046$ \\
 & $(7 \times 10^{-4} - 0.01)$ & $(3\times 10^{-4} - 0.02)$ & $(2\times 10^{-4} - 0.03)$ \\
\hline
\multirow{2}{*}{$f_{\rm d}^{(1)}$} & $0.75$ & $^{\star}8.1\times 10^{-3}$ & $^{\star}9.2 \times 10^{-4}$ \\
 & $(-)$ & $(7 \times 10^{-4} - 0.01)$ & $(5\times 10^{-4}-0.015)$ \\
\hline
$f_{\rm QCD}$ & $1.3$ & $0.14$ & $0.049$ \\
\hline
$f_{e\pm}$ & $4.4 \times 10^{-3}$ & $4.7 \times 10^{-4}$ & $1.6 \times 10^{-4}$ \\
\hline
\end{tabular}
\end{center}
\caption{
Characteristic frequencies associated with the cutoff by small-scale structure (cusps) ($f_*$) and the change of cosmology ($f_{\rm d}$) for $m_s = 10 {\rm TeV}$ with various values of $\phi_0$.
For $T_*$ we used \eq{eq:T-star} with $\xi=0.1$, $\beta_{\rm c}=1$, $w_s m_\phi = 1$, and $g_* = 75.75$ for $\alpha_*$.
For $T_{\rm d}$ we used \eq{eq:Td} with \eq{eq:Gamma-phi}, and took $\mu=m_\phi=m_s$, $m_A = 1.5 \mu$, and $B=(2/3) m_A$, and $g_*(T_{\rm d}) = 67$ as a roughly selected value which does not make a meaningful difference in our analysis.
In the first column, $f_{\rm d}^{(1)}$ and $f_{\rm d}^{(10)}$ stand for frequencies associated with $f_{\rm d}$ which may be probed if a detector is sensitive enough to probe $1$\% or $10$\% deviation, respectively \cite{Gouttenoire:2019kij}. 
These two frequencies appear due to the fact that it takes some time to reach to the scaling regime after a change of cosmology which causes a time-dependence of the loop formation efficiency.
Each of the frequency band in the parenthesis in some cells represents a rough LISA band sensitive to 1\% or 10\% deviation as a guide to see the possibility of detections.
The frequencies within the detection sensitivity band were marked with $\star$-mark at the front of the number in relevant cells. 
$f_{\rm QCD}$ and $f_{e\pm}$ stand for frequencies associated at the epochs of QCD phase transition and $e^+  e^-$ annihilations, respectively.
}
\label{tab:char-freq}
\end{table}

It should be noted that, even though we may get information about
the $B-L$ breaking scale,  $\phi_0$, and the SUSY scale,  $m_s$, from future experiments on gravitational waves,
 it is still non-trivial to identify the LSP dark matter.
The reason is as follows.
The amplitude of GW signals would provide information about a certain combination of $m_\phi$ and $\phi_0$ from their appearance in $\mu_s$ (see \eq{eq:mu-Nw}).
If the value $f_\ast$ is also extracted from the GW signals, it may be possible to fix the values of $m_\phi$ and $\phi_0$. 
However, in order to extract the precise value of $f_\ast$,   one needs to reduce the uncertainties associated with $w_s$,
the core width of strings, and the $N_w$ dependence of $f_\ast$, as can be seen from \eqs{eq:T-star}{eq:f-star}.
Furthermore,  as can be seen from 
 \eq{eq:mchiYchi}, 
 the relic density of neutralino dark matter 
has a model dependence on how and at which scale SUSY breaking is transmitted to the SM sector.
More precisely, it is determined mainly by the ratios,  $m_{\tilde \chi}/m_\phi$,  $|B|/m_A$ and $|\mu|/m_\phi$, where $T_{\rm d}$ depends on the latter two. 
Keeping such a model dependence in mind,  we have fixed the ratios to be certain values of the order of unity as an example.
This is simply to show that the neutralino LSP can account for the observed dark matter density without difficulty,  rather than to identify its nature upon the information of $m_s$ and $\phi_0$. 
One may scan over the full space of the model parameters to find the region where the neutralino LSP becomes the main component of the dark matter, and then specify the model further, such as the mediation mechanism of SUSY breaking, so as  to identify the nature of the neutralino LSP.
This is interesting,  but is out of the scope of this paper. 

If the high-frequency cutoff turns out to be $f_{\rm d}$, the situation is more tricky because of the degeneracy associated with
$T_{\rm d}$.
On the other hand, if the future data indicates a somewhat low $\phi_0$, axinos would become the most plausible candidate of the LSP dark matter.  

A comment is in order here.
If $U(1)_{B-L}$ were broken before the last primordial inflation, we do not expect the existence of strings in our Hubble patch.
Hence, a detection and discrimination of the characteristic spectrum of GW's shown in Figs.~\ref{fig:GWs-w-wo-zipper-ms3T} or~\ref{fig:GWs-w-zipper-ms10T} would indicate the presence of thermal inflation.

\section{Ultra high energy cosmic rays via $LH_u$ condensation}

It is well known that cosmic strings can produce ultra-high energy cosmic rays (UHECR's) dominantly from cusp-annihilations.
While the event rate is typically expected to be smaller than the observations by many orders of magnitude~\cite{Gill:1994ic,Blanco-Pillado:1998tyu}, 
 Ref.~\cite{Vachaspati:2009kq}  has pointed out that, when a light scalar field develops a large VEV inside the core of strings, the equation of motion of the light scalar field obtains a large source term (see also Ref.~\cite{Berezinsky:2011cp}).
Such a source term allows production of a large amount of highly boosted scalar particles, thereby enhancing the power of particle emission roughly by a factor of $\l( c_{\rm lin}^2/\mu_s \r) \sqrt{\phi_0/m}$ with $c_{\rm lin}$ being the dimensionful linear coupling of strings to the light scalar having a mass $m$. 
If the light scalar is the SM Higgs,  as an example, 
for $\phi_0 \sim \mathcal{O}(10^{13})\,{\rm GeV}$ with $c_{\rm lin}^2/\mu_s \sim \mathcal{O}(1)$, the cascade decays of booted SM Higgses may explain the observed UHECR's over 
the GZK limit~\cite{Vachaspati:2009kq}.

Our scenario naturally obtains the required amount of power-enhancement.
The core width of strings formed by the flat direction is roughly of the order of $m^{-1}$ with $m$ being the mass scale of the field and nothing but the scale of the soft SUSY breaking mass parameter,  i.e.~$m \sim m_s$.
The emission power of particles is then automatically enhanced by a factor of $\sqrt{\phi_0/m_s}$ as pointed out in Ref.~\cite{Cui:2007js}.
Since flaton decays dominantly to the SM Higgses, the enhancement factor in our scenario is already enough to explain observations without requiring any extra feature introduced in Refs.~\cite{Vachaspati:2009kq,Berezinsky:2011cp}.

There is another production channel of UHECRs in our scenario.
Although it was not intended, the mechanism discussed in Ref.~\cite{Vachaspati:2009kq} is realized accidentally but in a natural manner with one of the MSSM flat directions instead of 
the SM Higgs field.
As described in Section~\ref{sec:BAU}, in order to obtain enough amount of baryon   asymmetry after thermal inflation, 
one can rely on a late-time Affleck-Dine leptogenesis along the $LH_u$ flat direction.
It works well because the effective mass-squared of the flat direction depends on the $U(1)_{B-L}$ breaking field $\phi$ such that it becomes positive only if $\phi$ is close to its VEV, $\phi_0$.
Such a possibility can arise easily and naturally by the interplay of SUSY breaking parameters.
In regard of cosmic strings, this means that in the region of the core of strings where $\phi < \phi_0$,   the $LH_u$ flat direction becomes unstable and  develops a non-zero VEV.    

The expected direct flux at Earth due to the presence of $LH_u$ condensation can be estimated easily by modifying Eq.~(53) in Ref.~\cite{Vachaspati:2009kq} with the following replacement:
\beq
M^2 \ln \l( \frac{M}{m} \r) \to \l( \pi w_s^2 |m_{LH_u}^2(0)| \phi_{\rm AD, in} \r)^2 \ln \l( \frac{w_s^{-1}}{m} \r),
\eeq
where $\phi_{\rm AD, in}$ is the VEV of the $LH_u$ flat direction inside of the core.
The resultant flux reads
\beq \label{eq:p-flux-formula-1}
k \frac{d \Phi}{d k} 
= 24 \pi^3 c_\ell w_s^2 \l| m_{LH_u}^2(0) \r| \ln \l( \frac{1}{m w_s} \r) \frac{\phi_{\rm AD, in}^2}{\l( \Gamma_{\rm ptl} G \mu_s \r)^2 k^2} \frac{R^3}{t^4},
\eeq
where $c_\ell = 10$ \footnote{Here $c_\ell$ stands for $A$ in Ref.~\cite{Vachaspati:2009kq}. It is a coefficient related to the population of string loops.}, and $\Gamma_{\rm ptl} = \mathcal{O}(1)$ is the coefficient associated with the power for particle emissions.
The flux per unit area on earth is then found as \footnote{It should be noted that the string tension $\mu_s$ is smaller than the case of Ref.~\cite{Vachaspati:2009kq} by about an order of magnitude for a given symmetry breaking scale.}
\bea \label{eq:p-flux-formula-2}
k \frac{d \Phi}{dA dk} 
&\simeq& \frac{1.4 \times 10^{-4} \l( m_\phi w_s \r)^2}{{\rm km}^2 \cdot {\rm yr} \cdot {\rm sr}} \frac{\l| m_{LH_u}^2(0) \r|}{m_\phi^2} 
\nonumber \\
&& \times
\l( \frac{\phi_{\rm AD, in}}{10^{11} {\rm GeV}} \r)^2 
\l( \frac{10^{13} {\rm GeV}}{\phi_0} \r)^2
\l( \frac{10^{11} {\rm GeV}}{k} \r)^2
\l( \frac{R}{15 {\rm Mpc}} \r)^3,
\eea
using $\ln \l(m w_s \r)=-1$ as an example, 
where 
the surface area of the Earth, $5.1 \times 10^9 {\rm km}^2$, $m \sim m_\phi \sim m_s = 10^4\, {\rm GeV}$, and the attenuation length of a energetic proton was taken to be $R_p \approx 15 {\rm Mpc}$ at $k = \mathcal{O}(10^{11}) {\rm GeV}$ \cite{Kotera:2011cp,Dova:2015hpq}.
The observed flux of UHECR's at $k \sim 10^{11} {\rm GeV}$ is about $10^{-3}/{\rm km}^2 \cdot {\rm yr} \cdot {\rm sr}$ \cite{TelescopeArray:2021zox}.
For the decay of the $H_u$ field, this estimation may be directly applicable.

If kinematically allowed, thanks to gauge interactions, the sneutrino field ${\tilde \nu}$, the other component of the $LH_u$ flat direction and the superpartner of the associated SM neutrino field, decays primarily neutrinos and neutralinos, 
\beq
{\tilde \nu}_\alpha \to \nu_\alpha + {\tilde \chi}
\eeq
with $\alpha$ being the flavor index.
If the axino is the LSP, there will be channels of 
\beq
{\tilde \chi} \to q_\alpha +\bar{q}_\alpha + {\tilde a}
\eeq
with $q_\alpha$ being a quark field.
This process is due to radiatively generated axino-quark-squark interactions \cite{Covi:2002vw}.
Extremely energetic neutrinos can be produced in such ways, travelling a horizon size distance.
Such neutrinos might be responsible for the observed UHECR's over the GZK limit with the help of a large enhancement of the flux in \eq{eq:p-flux-formula-2} (see
Ref.~\cite{Berezinsky:2011cp} for an estimation of the neutrino flux).
Also, there will be highly boosted LSPs (either sneutrino itself, neutralino, or axino).
The whole detail of those ultra-high-energy particles is out of the scope of this work, and will be discussed elsewhere.

\section{Conclusions}

In this work, we have proposed a local $U(1)_{B-L}$ extension of the MSSM, which is characterized by a superpotential consisting of all gauge-invariant operators up to leading higher order ones involving the MSSM fields, right-handed neutrinos,  and two oppositely charged ${B-L}$ Higgses.

Our model can realize a period of thermal inflation as already shown in a similar model considered in Ref.~\cite{Jeannerot:1998qm}.
It turns out that resolution of the cosmological moduli problem requires $m_s \gtrsim \mathcal{O}(10)\, {\rm TeV}$ and $\phi_0 = \mathcal{O}(10^{12-15})\, {\rm GeV}$ if the initial oscillation amplitude of moduli is around the Planck scale as naturally expected. 
The baryon asymmetry of the universe can be generated successfully by the late-time Affleck-Dine leptogenesis working right after thermal inflation.
It requires one of the mass parameters for the active neutrinos to be several orders of magnitude smaller than $\mathcal{O}(10^{-2})\,{\rm eV}$ in addition to the condition, $m_L^2 + m_{H_u}^2 < 0$ with $m_i^2$ being the soft SUSY breaking mass squared parameters.
Dark matter in our scenario can be either sneutrino, the lightest MSSM neutralino, or KSVZ-type axino even if $m_s = \mathcal{O}(10)\,{\rm TeV}$,  while the  QCD axion solving the strong CP problem is still another good dark matter candidate.

As well known,  $U(1)_{B-L}$  breaking  leads to  the formation of cosmic strings.
Stochastic gravitational wave backgrounds (SGWB's) are produced dominantly from cosmic string loops.
In our scenario, the constraint from PPTA data on gravitational waves is interpreted as a bound on the symmetry breaking scale as $\phi_0 \lesssim 3 \times 10^{13}\, {\rm GeV}$, but the expected signal can be within the reach of near-future experiments such as SKA, LISA, and DECIGO, depending on the model parameter space. 
  
Because of the flatness of the associated potential, in contrast to the conventional Abelian-Higgs model, strings in our scenario are very thick, i.e.~the mass scale of the Higgs field is many orders of magnitude smaller than that of the gauge field, $m_V / m_\phi \sim g_{BL} \phi_0/m_s= \mathcal{O}(10^{8 - 11})$ with $g_{BL} = \mathcal{O}(0.1-1)$.
It has been noticed that such a property causes, so-called,  the zippering effect.
Encoding the effect in the emission power of GWs averaged over allowed winding numbers at a given epoch, we showed that a sizable spectral deviation is expected relative to the GWs in the conventional Abelian-Higgs model.
Also, the expected GW signals have information of either the mass scale or the decay temperature of the flaton field, which appears as a bending point at $\mathcal{O}(10^{-3}-1)\, {\rm Hz}$.
As shown in Fig.~\ref{fig:Probing-TI} with Table~\ref{tab:char-freq} in the body of the paper, including the bending feature, the spectral deviation is large enough and can be probed at forthcoming experiments such as LISA or DECIGO.

It should be noted that, if all the gauge couplings are to be unified, 
the strength of gauge interaction is not very different from order unity, and the existence of a gauge-charged flat-direction can be regarded as a characteristic nature of supersymmetric theories.
The clue of flat-direction in the signal of GW's in our scenario may then be interpreted as a clue of supersymmetry and the presence of thermal inflation.

As the final interesting point, as pointed out in Ref.~\cite{Cui:2007js}, thanks to the flatness of the symmetry breaking potential leading to a very wide core of strings, the particle emission from cusp-annihilations of string loops is enhanced by many orders of magnitude.
Interestingly, we found that the observed flux of ultra high energy cosmic rays over the GZK limit can be naturally obtained from the decay of ${B-L}$ Higgses to SM Higgses as the dominant channel.
We have also found that accidentally but inevitably there appears a condensation of the $LH_u$ flat direction inside of the core of stings, and this can be the main source of the observed ultra-high energy cosmic rays, especially accompanied by highly boosted supersymmetric particles,
for instanace,  sneutrino, neutralino, or axino.

\medskip


\medskip


\section*{Acknowledgments}

\noindent
This work was supported by the National Research Foundation of Korea grants by the Korea government: 2017R1D1A1B06035959 (W.I.P.),  2022R1A4A5030362 (W.I.P.),   2021R1A4A5031460 (K.S.J.),
and RS-2023-00249330 (K.S.J.).
It was also supported by the Spanish grants PID2020-113775GB-I00 (AEI/10.13039/501100011033) and CIPROM/2021/054 (Generalitat Valenciana) (W.I.P.).

\appendix

\section{Deviation from the $D$-flat direction}
\label{sec:app-dev-flat}
For the superpotential term,
\beq
W \supset \frac{\lambda_\Phi}{2M} \l( \Phi_1 \Phi_2 \r)^2,
\eeq
the scalar potential of $\phi_{1,2}$ is found as
\beq \label{eq:V-tree}
V = V_0 + V_{\rm soft} + V_F + V_D,
\eeq
with each term given by
\bea
V_{\rm soft} &=&  m_1^2 |\phi_1|^2 + m_2^2 |\phi_2|^2 - \l[ B_\Phi \mu_\Phi \phi_1 \phi_2 + \frac{ A_\Phi \lambda_\Phi}{2M} \l( \phi_1 \phi_2 \r)^2 + {\rm c.c.} \r],
\\
V_F &=& \l| \mu_\Phi + \frac{\lambda_\Phi}{M} \phi_1 \phi_2 \r|^2 \l( |\phi_1|^2 + |\phi_2|^2 \r),
\\
V_D &=& \frac{1}{2} g_{BL}^2 \l( |\phi_1|^2  - |\phi_2|^2 \r)^2,
\eea
where $g_{BL}$ denotes the $U(1)_{B-L}$ gauge coupling.
Hence, the scalar potential of $\phi_{1,2}$ is
\bea
V
&=& V_0 + \l( m_1^2 + |\mu_\Phi|^2 \r) |\phi_1|^2 + \l( m_2^2 + |\mu_\Phi|^2 \r) |\phi_2|^2 
\nonumber \\
&-& \l[ B_\Phi \mu_\Phi \phi_1 \phi_2 + \frac{ A_\Phi \lambda_\Phi}{2M} \l( \phi_1 \phi_2 \r)^2 - \frac{\mu_\Phi \lambda_\Phi \l( |\phi_1|^2 + |\phi_2|^2 \r)}{M} \phi_1\phi_2 + {\rm c.c.} \r]
\\
&+& \l|\frac{\lambda_\Phi}{M} \phi_1 \phi_2 \r|^2 \l( |\phi_1|^2 + |\phi_2|^2 \r) + \frac{1}{2} g_{BL}^2 \l( |\phi_1|^2  - |\phi_2|^2 \r)^2.
\eea
We assume  
\beq
m_{1,2}^2 + |\mu_\Phi|^2 < 0,
\eeq
from the scale of $m_s$ up to at least some intermediate scales.
We also assume ${\rm Arg}\l[B_\Phi \mu_\Phi \r] = {\rm Arg} \l[A_\Phi \lambda_\Phi \r] = 0$, and ${\rm Arg}\l[\mu_\phi \lambda_\Phi \r]=\pi$ for simplicity.
Then, along the direction with ${\rm Arg}\l[ \phi_1 \phi_2 \r] =0$ where the true vacuum is 
determined by 
\bea
\partial_1 V 
&=& \l[ 2\l( m_1^2 + |\mu_\Phi|^2 \r) + \frac{\lambda_\Phi^2}{M^2} \l( 4 |\phi_1 \phi_2|^2 + 2 |\phi_2|^4 \r) + 2  g_{BL}^2 \l( |\phi_1|^2 - |\phi_2|^2 \r)  \r] |\phi_1|
\nonumber \\
&& - 2 \l[ B_\Phi \mu_\Phi  + \frac{A_\Phi \lambda_\Phi}{M} |\phi_1 \phi_2| + \frac{\mu_\Phi \lambda_\Phi}{M} \l( 3 |\phi_1|^2 + |\phi_2|^2 \r) \r] |\phi_2|,
\\
\partial_2 V 
&=& \l[ 2\l(m_2^2 + |\mu_\Phi|^2 \r) + \frac{\lambda_\Phi^2}{M^2} \l( 4 |\phi_1 \phi_2|^2 + 2 |\phi_1|^4 \r) - 2  g_{BL}^2 \l( |\phi_1|^2 - |\phi_2|^2 \r)  \r] |\phi_2|
\nonumber \\
&& - 2 \l[ B_\Phi \mu_\Phi  + \frac{A_\Phi \lambda_\Phi}{M} |\phi_1 \phi_2| + \frac{\mu_\Phi \lambda_\Phi}{M} \l( 3 |\phi_2|^2 + |\phi_1|^2 \r) \r] |\phi_1|.
\eea
At the true vacuum, requiring $\partial_1 V = \partial_2 V = 0$ and setting $\phi_1 = \phi_2 + \Delta \phi \equiv \phi/\sqrt{2}$ with $|\Delta \phi| \lll |\phi|$, 
one finds
\bea
\Delta \phi 
&=& \frac{\l( m_1^2 - m_2^2 \r) |\phi|/\sqrt{2}}{-\l( m_2^2 + |\mu_\Phi|^2 \r) + B_\Phi \mu_\Phi + 2 g_{BL}^2 |\phi|^2 + A_\Phi \lambda_\Phi |\phi|^2/2M - \lambda_\Phi^2 |\phi|^4/4M^2}
\\
&\simeq& \frac{\l( m_1^2-m_2^2 \r)}{2\sqrt{2}g_{BL}^2 |\phi|}.
\eea
Hence, unless $g_{BL}$ is smaller than unity by many orders of magnitude, $\Delta \phi$ is negligible. 
Therefore, ignoring the $D$-term contribution,  
the scalar potential should satisfy
\beq
0 = - \overline{m^2} - \frac{A_\Phi \lambda_\Phi}{2M} |\phi|^2 + \frac{3\lambda_\Phi^2}{4M^2} |\phi|^4,
\eeq
at the vacuum, 
where we have also ignored $\mu_\Phi$-dependent terms,
 and $\overline{m^2} \equiv - \l( m_1^2 + m_2^2 \r) /2$, leading to
\beq
|\phi_0|^2 = \frac{A_\Phi M}{3 \lambda_\Phi} \l( 1 + \sqrt{1 + \frac{12 \overline{m^2}}{A_\Phi^2}} \r).
\eeq

\section{Annihilation cross sections of various neutralinos}
\label{sec:app-ann}
The thermal-averaged annihilation cross sections of various neutralino  LSP are given by \cite{Griest:1988ma,Olive:1989jg,Olive:1990qm,Gondolo:1990dk}
\begin{itemize}
\item {\bf Higgsino LSP}

This would be the case in high scale SUSY breaking.
The higgsino LSP has
\bea
\langle \sigma v_{\rm rel} \rangle
\simeq \frac{\pi \alpha_2^2}{m_{\tilde h}^2} \l( \frac{1}{2} + \frac{1}{4 \cos^4 \theta_W} \r) \simeq 10^{-9} {\rm GeV}^{-2} \l( \frac{1 {\rm TeV}}{m_{\tilde h}} \r)^2.
\eea
Thus,  one finds that the decay temperature of the flaton be around $10$~GeV.
\item {\bf Bino LSP}

For  the bino LPS,  including the leading order temperature-dependence, one obtains 
\beq
\langle \sigma v_{\rm rel} \rangle \simeq \frac{32 \pi}{27} \alpha_1^2 \frac{m_{\tilde B}^2}{\l( m_{\tilde{t}_R}^2 + m_{\tilde{B}}^2 - m_t^2 \r)^2} \l( 1 - \frac{m_t^2}{m_{\tilde{B}}^2} \r)^{1/2} \l( \frac{m_t^2}{m_{\tilde B}^2} + 2 \frac{T}{m_{\tilde B}} \r).
\eeq
In this case, one needs lower $T_{\rm d}$ due to a smaller annihilation cross section compared to the higgsino LSP case.
\item {\bf Wino LSP}

For  the wino LPS, one finds
\beq
\langle \sigma v_{\rm rel} \rangle \simeq \frac{8\pi \alpha_2^2 \l( 1 - x_W \r)^{3/2}}{m_{\tilde{W}}^2 \l( 2 - x_W \r)^2},
\eeq
where $x_W = m_W^2/m_{\tilde W}^2$.
The wino-LSP is likely to be underproduced even for $m_{\tilde{t}} \sim m_s = \mathcal{O}(10) {\rm TeV}$.

\item {\bf Sneutrino LSP}

For the sneutrino LSP, the thermal-averaged annihilation cross-section can be found as \cite{Hagelin:1984wv} (see also Ref.~\cite{Falk:1994es})
\beq
\langle \sigma v_{\rm rel} \rangle \simeq \pi \sum_{i}^4 \frac{\alpha_2^2}{m_{{\tilde \chi}_i}} \l( c_{i1} - c_{i2} \tan \theta_W \r)^4,
\eeq
where $m_{{\tilde \chi}_i}$ is the mass of the neutralino mass eigenstate $i$, $c_{i1}$ and $c_{i2}$ are respectively the wino and bino components in the mass eigenstate, and $\theta_W$ is the Weinberg angle.
\end{itemize}


\end{document}